\documentstyle[12pt]{article}
\newcommand{\nc}{\newcommand}
\nc{\Rerm}{{\rm Re}} \nc{\Imrm}{{\rm Im}} \nc{\Rrm}{{\rm R}}
\nc{\Trm}{{\rm T}} \nc{\Drm}{{\rm D}} \nc{\Crm}{{\rm C}}
\nc{\Brm}{{\rm B}} \nc{\Lrm}{{\rm La}} \nc{\Srm}{{\rm S}}
\nc{\Irm}{{\rm I}} \nc{\IIrm}{{\rm II}} \nc{\IIIrm}{{\rm III}}
\nc{\grm}{{\rm g}} \nc{\Inrm}{{\rm in}} \nc{\BLrm}{{\rm BL}}
\nc{\inrm}{{\rm in}} \nc{\pirm}{{\rm p}} \nc{\frm}{{\rm f}}
\nc{\irm}{{\rm i}} \nc{\clrm}{{\rm cl}} \nc{\drm}{{\rm d}}
\nc{\Rcal}{{\cal R}}
\nc{\noi}{\noindent} \nc{\pa}{\partial}
\nc{\vecna}{\mbox{\boldmath $\nabla$}} \nc{\xbf}{\mbox{\boldmath $x$}}
\nc{\imp}{\mbox{\boldmath $p$}} \nc{\vbf}{\mbox{\boldmath $v$}}
\nc{\Bbf}{\mbox{\boldmath $B$}} \nc{\rbf}{\mbox{\boldmath $r$}}
\nc{\lan}{\langle} \nc{\ran}{\rangle}

\begin{document}

\centerline{\large\bf Tunnelling Times:}
\centerline{\large\bf An Elementary Introduction$^{(\dagger)}$}

\footnotetext{$^{(\dagger)}$ Work partially supported by INFN and MURST
(Italy); and by I.N.R., Kiev, and S.K.S.T. (Ukraine).}

\begin{center}
\vspace*{1cm} {Giuseppe Privitera\break {\em Facolt\`a di Ingegneria,
Universit\`a di Catania, Catania, Italy;\break e-mail:
{\rm giuseppe.privitera@dmfci.unict.it}}}\\

\

{Giovanni Salesi\break {\em Facolt\`a di Ingegneria, Universit\`a di
Bergamo, Dalmine (BG), Italy;\break  {\rm and} I.N.F.N.--Sezione di Milano,
Milan, Italy.}}\\

\

{Vladislav S. Olkhovsky\break {\em Institute for Nuclear Research,
Kiev-252028; Research Center "Vidhuk", Kiev, Ukraine.}}\\

\

{Erasmo Recami\break {\em Facolt\`a di Ingegneria, Universit\`a di Bergamo,
Dalmine (BG), Italy;\break {\rm and} I.N.F.N.--Sezione di Milano, Milan,
Italy;
\break e-mail: {\rm recami@mi.infn.it} }}
\end{center}
\

\

\begin{abstract}
In this paper we examine critically and in detail some existing
definitions for the tunnelling times, namely: the phase-time; the
centroid-based times; the B\"{u}ttiker and Landauer times; the Larmor times;
the complex (path-integral and Bohm) times; the dwell time, and finally the
generalized (Olkhovsky and Recami) dwell time, by adding also some numerical
evaluations. \ Then, we pass to examine the equivalence between quantum
tunnelling and ``photon tunnelling" (evanescent waves propagation), with
particular attention to tunnelling with Superluminal group-velocities
(``Hartman effect"). \ At last, in an Appendix, we add a bird-eye view of
all the experimental sectors of physics in which Superluminal motions seem to
appear.
\end{abstract}

\

\baselineskip 0.7cm

\section{Introduction} 

Let us consider a freely moving particle which, at a given time, meets a
potential barrier higher than its energy. As it is known, quantum
mechanics implies a non-vanishing probability for the particle
to cross the
barrier (i.e., the {\em tunnel effect\/}). We may actually ask ourselves if it is
possible to define a {\em time duration}, and therefore an average
{\em speed}, for the barrier
crossing; in that case we might try to calculate and measure those quantities.
It seems surprising that no answer to such a question, so (apparently)
straightforward, has gained general acceptance yet. The problem was
first pointed out in 1931 by Condon[1]; an early attempt to solve it
was due to MacColl[2] a year later. Afterwards, this subject was
almost ignored up to the fifties, apart from the introduction of a
quantum-theoretical observable for Time in scattering theory
(For a simple introduction in quantum mechanics
of a non-selfadjoint {\em but hermitian} time-operator, see refs.[3]).
However, the interest on the subject has grown up, mostly during the
last twentyfive years, for the increasing use of high-speed electronic
devices (based on tunnelling processes) as well as for the acknowledged
importance of the tunnel effect in nuclear fission and fusion below threshold.
As a consequence, after 1987, several[4-6]
theoretical reviews about tunnelling times appeared in the literature.
Unfortunately on the experimental side it has always been rather difficult to
perform measurements on particles, such as electrons, in order to check
the theoretical predictions. Only in recent years various
measurements of the sub-barrier ``transmission times" (for microwaves and optical
photons) have been performed, by exploiting the mathematical equivalence
between the behaviour of (classical, relativistic) ``evanescent''
electromagnetic waves[7] and the (quantum, non-relativistic) tunnelling of a
particle (or photon). We shall come back to the mentioned mathematical
equivalence when dealing with the relevant experiments.

Tunnel effects can be
met in various physical processes, as scattering, disexcitation of metastable
states, fission and fusion above threshold, and so on.  Anyway, we shall
analyze only the one-dimensional tunnelling of a particle through a constant
potential barrier$^{\#1}$ $V_{0}$, localized in the interval $[0,d]$ (see
Fig.1): a case that represents well enough the currently employed
experimental setups. \footnotetext{$^{\#1}$ Only in the case of the
B\"{u}ttiker-Landauer time, the barrier will be a function of time too.}
Despite the large quantity of theoretical papers on the subject, a
universally accepted approach to tunnelling time does not seem to exist,
till now.  We shall therefore group the existing approaches into four
classes.\\

In the first group we consider all times built by ``following'' the
entering packet while it crosses the barrier. \
After the choice of a particular feature of the packet, e.g., its
central peak, one can compare the incoming and the outgoing peak, obtaining
their time correlation. Various authors, alternatively,
have considered the ``centroid'' (that is, the wavepacket center-of-mass),
or the sharp wave-front of a ``step-like'' wave. To this class of times
it belongs the {\em phase time}, which is obtained, on employing a
definition of group-velocity, through the ``stationary-phase
approximation''. It appears to agree rather well with the so far
available experimental data.
Among the more common critical remarks to this approach, let us recall the following: The outgoing
peak might not always correspond to the incoming one, because of the
presence, among
the Fourier components of the wavepacket, of frequencies close or above the
barrier energy. These high frequencies would reach the barrier earlier,
since they travel towards the barrier less distorted by the reflected
waves with respect to the low-frequency components.
As a consequence, besides the dispersion effects (often present also while
approaching the barrier), there appears an effective {\em acceleration}
of the wavepacket.  The further known effects taking place in the barrier
region lead to the conclusion that the packet may even seem to exit the barrier
before the main peak has entered it.\hfill\break

A second approach consists in assuming the existence of some degrees of
freedom in the barrier-particle system, in order to be able to define an
``internal clock'' which yield the time spent by the particle inside the
barrier. Through the
effect of the ``barrier clock'' on the particle or, viceversa, through the
effect of the barrier on the ``particle clock'', we infer the duration of
their interaction during the barrier crossing. B\"{u}ttiker and
Landauer, for example, tried to deduce the tunnelling time by calculating the
energy quanta exchanged by a particle crossing a square barrier endowed with a
time-varying height ({\em B\"{u}ttiker-Landauer time}). Another example can be
the
measurement of the electron spin ``flip'' when inside the barrier a uniform
magnetic field is present ({\em Larmor times}). This ``clock
approach'' allows a wide choice in the degrees of freedom of the system used
as a clock; furthermore, it yields some experimental procedures in order to
check the theoretical predictions. Nevertheless, it has been criticized by some
authors, since not all clocks are equivalent, and, even more, since such
clocks,
besides changing the number of the degrees of freedom of the system,
involve invasive processes, which affects the experimental outputs. We cannot
be sure that the duration of the mentioned interaction (during which the
state of the system suffers a perturbation, anyway) actually coincides
with the crossing time or with the reflection time.\hfill\break

The third approach attributes to the below-barrier motion of a particle a set
of ``semiclassical'' trajectories, with respect to which an average tunnelling time
can be evaluated. The paths can be built up in various ways, e.g., through the
{\em Feynman path-integrals}, the {\em Bohm mechanics}, or the {\em Wigner
distribution}. Of course, each of these methods carries a distribution of
crossing times. One of the inconveniences of this approach is the complex
nature of the computed time quantities. Nevertheless, we can extract from these
times some real quantities ---the magnitude, the real part, the imaginary
part--- which result to be strictly related with some of the above definitions of
tunnelling times. Just for this reason the physical interpretation of the
semiclassical times appears rather interesting.\hfill\break

The last class of tunnelling times we are going to consider is the one which
starts from the definition of {\em dwell time}. Such a quantity is defined as
follows:
$$
\tau^\Drm(x_1,x_2;k)=j_\inrm^{-1} \int_{x_1}^{x_2}|\psi (x,k)|^2 \drm x, \eqno(1.1)
$$
that is, it is the ratio between the probability density in the tunnelling region
and the incoming flux entering the barrier $j_{\inrm}$. The inconvenience of
this definition is that it actually yields the dwell time inside the
barrier, but without distinguishing between the transmission and the
reflection channels. In this regard, an equation ---often considered obvious
even if not universally accepted--- expected to link the times of the above
two channels is the following:
$$
\tau^\Drm=|T(k)|^2\tau_\Trm + |R(k)|^2\tau_\Rrm. \eqno(1.2)
$$
This equation, even if correct, seems to be not sufficient to determine
uniquely $\tau_\Trm$ and $\tau_\Rrm$. To overcome this difficulty, some authors
introduced the so-called ``space approaches''. Those pictures link the
time spent inside the barrier with the reflection and transmission times, and
are built up through the standard quantum probabilistic interpretation of the
reflection and transmission ``currents''.\\

Let us anticipate that some definitions of tunnelling time we are going
to analyze are sometimes regarded to be more suitable than others: in fact,
they do not involve the so-called {\em Hartman effect} (see Sect.12)
and do agree with eq.(1.2). Nevertheless, they result to disagree
with the experimental data.  We shall actually see that the really (physically)
reasonable definitions of tunnelling time do all imply, among the others,
he existence of Hartman-type effects, which have been confirmed by
experiments.

This review is meant to be an elementary one; for further details
see ref.[8].

\

\section{Assumptions and notations} 

Let us suppose to have solved exactly the stationary case. Then, for each
given energy \ $E=\hbar ^{2}k^{2}/2m$, \ we have:
$$
\psi (x;k)=
\cases{\psi_\Irm= e^{ikx}+R(k)e^{-i(kx-\beta)} & $x\leq 0$\cr
                  \psi_{\IIrm}=\chi (x;k) & $0\leq x\leq d$\cr
                  \psi_{\IIIrm}=T(k)e^{i(kx+\alpha)} &$ x\geq d$\cr}\eqno
(2.1)
$$
where $R(k)$ and $T(k)$ are the reflection and transmission amplitudes,
respectively, so that
\[
R(k)=\sqrt{1-T(k)^{2}}\,,
\]
quantities \ $\beta =\beta (k)$, $\alpha =\alpha (k)$ \ being the
respective phase delays. In the particular case of a square barrier we
have
\[
V(x)=
\cases{V_0 & $0\leq x\leq d$\cr
                    ${\rm 0}$ & {\rm elsewhere},\cr }
\]
while \ $\chi (x;k),\;R(k),\;T(k),\;\alpha (k),\;\beta (k)\;$ are
analytically known. Therefore, we shall write
$$
\chi (x;k)=
\cases{A(k)e^{-\kappa x}+B(k)e^{\kappa x} & $E<V_0$\cr
A(k)e^{-i\kappa x}+B(k)e^{i\kappa x} & $E > V_0$\cr},\eqno(2.2)
$$
\noindent with
\[
\kappa =
\cases{\sqrt{2m(V_0-E)}/\hbar,&$E<V_0$\cr
                             \sqrt{2m(E-V_0)}/\hbar,&$E > V_0$\cr}
\]
$$
R(k)=
\cases{\displaystyle{(k^2-\kappa^2)\sinh (\kappa d)\over
[4k^2\kappa^2+(k^2+\kappa^2)^2\sinh^2(\kappa d)]^{1\over2}}&
$E<V_0$\cr\displaystyle {(k^2+\kappa^2)\sin (\kappa d)\over
[4k^2\kappa^2+(k^2-\kappa^2)^2\sin^2(\kappa d)]^{1\over2}}&
$E > V_0$\cr}\eqno(2.3)
$$
$$
T(k)=
\cases{\displaystyle{2k\kappa\over
[4k^2\kappa^2+(k^2+\kappa^2)^2\sinh^2(\kappa d)]^{1\over2}}&
$E<V_0$\cr\displaystyle {2k\kappa\over
[4k^2\kappa^2 +(k^2-\kappa^2)^2\sin^2(\kappa d)]^{1\over2}}&
$E > V_0$\cr}\eqno(2.4)
$$
$$
\beta (k)=
\cases{\arctan\left(\displaystyle{-2k\kappa\over(k^2-\kappa^2)}
\coth (\kappa d)\right),&$E<V_0$\cr
\arctan\left(\displaystyle{-2k\kappa\over(k^2+\kappa^2)}
\cot (\kappa d)\right), &$E > V_0$\cr}\eqno(2.5)
$$
$$
\alpha (k)=
\cases{\arctan\left(\displaystyle{(k^2-\kappa^2)\over 2k\kappa}
\tanh (\kappa d)\right),&$E<V_0$\cr
\arctan\left(\displaystyle{(k^2+\kappa^2)\over 2k\kappa}
\tan (\kappa d)\right), &$E > V_0$\,.\cr}\eqno(2.6)
$$
In the following we shall not consider simple stationary waves, but
the wavepackets 
$$
\Psi (x;t)\sim\int\drm k\,f(k-k_{0})\;\psi(x;k)\,e^{-i{\frac{E(k)t}{\hbar}}}
=\int \drm E\ g(E-E_{0})\;\psi (x;k(E))\,e^{-i{\frac{Et}{\hbar}}}.\eqno(2.7)
$$
We get also:
$$
\rho =|\psi (x)|^{2}\qquad \qquad j=\Rerm\,\left\{ {\frac{i\hbar}{2m}}\psi
(x){\frac{\partial}{\partial x}}\psi (x)\right\} \,.\eqno(2.8)
$$
In the electromagnetic framework (we always suppose photons and
microwaves in the T.E. or T.M. modes), $\psi $ represents the scalar part of
one of the two components of the field.

Finally, let us define the {\em equivalent time}, $\tau_{{\rm eq}}^\Trm$,
namely the time which the particle would spend in crossing the barrier
region {\em in the absence of the barrier}: i.e., \ $\tau^\Trm_{{\rm eq}}=md/\hbar k$.

\

\noindent Let us now review the definitions of some above-mentioned times.

\

\section{Phase time} 

Let us consider a very narrow wavepacket around a wave-number $k_0$.
The picture of its time evolution is often very difficult
because of its own dispersive nature. Anyway, under suitable conditions, it
is possible to follow the position of the peak of a {\em symmetric} packet
with a good precision, neglecting the dispersion effects[5,8]. We shall try therefore
to identify the packet taking the peak as reference point. To this end we
shall use the method of the stationary phase.

The peak of the packet is formed by those Fourier components for which
the phase variation in the surrounding of $k_{0}$ is reduced enough, so much
that they do not interfere destructively. Also the transmitted and reflected
packets will be described by wave-functions, corresponding to a small range
of frequencies:
\[
\psi(x;k) \sim {\rm exp}\,\left\{i\left[kx-\frac{E(k)t}{\hbar} +
\alpha(k)\right]\right\}\,.
\]
If we want to follow the position $x_{\pirm}(t)$ of the peak, we must look
for which values of $x_{\pirm}(t)$ the phase is stationary
(maximum) at a given$^{\#2}$ time $t$.
\footnotetext{$^{\#2}$Applying the same reasoning to the incident part, we'll
have: $x_{\pirm}(t)=\hbar^{-1}(\drm E/\drm k)\,t = (\drm\omega/\drm k)\,t$,
where $(\drm\omega / \drm k)$ is just, from its own definition, the
group-velocity of the incoming packet. Moreover we notice that
$v_{\grm}=(\drm\omega /\drm k)$ will depend on the dispersion relation $\omega
(k)$ of the medium in which the packet propagates.} Then we must have:
$$
{\frac{\drm}{\drm k}}\left( kx_{\pirm}(t)-{\frac{E(k)t}{\hbar}}+\alpha
(k)\right) =0\qquad \Rightarrow \qquad x_{\pirm}(t)={\frac{1}{\hbar}}{\frac{
\drm E}{\drm k}}t-{\frac{\drm\alpha}{\drm k}}\,.\eqno(3.1)
$$
Quantity $\alpha ^{\prime}(k_{0})=(\drm\alpha /\drm k)_{k_{0}}$ represents
the space delay $\delta x$ caused by the tunnelling process. Dividing by $v_{
\grm}$ (group-velocity of the wavepacket) we obtain the time delay
$$
\delta \tau _{\Trm}={\frac{\delta x}{v_{\grm}}}=(v_{\grm})^{-1}{\frac{\drm
\alpha}{\drm k}}=\left( {\frac{1}{\hbar}}{\frac{\drm E}{\drm k}}\right)
^{-1}{\frac{\drm\alpha}{\drm k}}=\hbar {\frac{\drm\alpha}{\drm E}}.\eqno
(3.2)
$$
Notice that we are computing $v_{\grm}$ and the other quantities
at $k=k_{0}$ (we'll see later on whether and when this procedure may be
considered correct).

For reasons which we shall later explain, we define as {\em phase time} the
total time $\tau _{\Trm}^{\varphi}(x_{1},x_{2};k)$, spent by a particle
between two points, $x_{1}$ and $x_{2}$, external to the barrier
and sufficiently distant from it, i.e., such that $x_{1}\ll 0$ and
$x_{2}\gg d$. Then
$$
\tau _{\Trm}^{\varphi}(x_{1},x_{2};k)={\frac{1}{v_{\grm}}}
(x_{2}-x_{1}+\alpha ^{\prime}(k))\,,\eqno
(3.3)
$$
and analogously, for the reflected particles,
$$
\tau _{\Rrm}^{\varphi}(x_{1},x_{2};k)={\frac{1}{v_{\grm}}}(-2x_{1}+\beta
^{\prime}(k))\,.\eqno(3.4)
$$
Since both $\tau _{\Trm}^{\varphi}$ and $\tau _{\Rrm}^{\varphi}$ depend
linearly on $x_{1}$ and $x_{2}$, we might try to extrapolate the crossing
and reflection times directly from them: by letting $x_{1}$ and $x_{2}$
approach $0$ and $d$,
respectively. Yet, this is not correct. In fact, since we have supposed the
components of the packet tightly distributed around the wave-number $k_{0}$,
the space spreading of the packet is the order of
$\sigma ^{-1}$ ($\sigma\equiv\Delta k$).
Therefore the packet will be the more spread, the more it is 
peaked around $k_{0}$. Thus the incident part and the reflected part of the
wave-function will be able to interfere with each other also at a certain
distance ($\sim \sigma ^{-1}$) from the barrier (see Fig.2).
Moreover, since we are employing a stationary approximation, we do not
really follow the packet in its time evolution, but only observe
(asymptotically) the phase-delay corresponding to the wave-number $
k_{0}$. Let us define
$$
\Delta\tau_\Trm^\varphi ={\frac{1}{v_\grm}}[d+\alpha^\prime (k)],\eqno(3.5)
$$
$$
\Delta\tau_\Rrm^\varphi ={\frac{1}{v_\grm}}[\beta^\prime (k)],\eqno(3.6)
$$
that we call {\em extrapolated phase times}.  We have always to keep in
mind the purely asymptotic character of such definitions. For a
square barrier we have:
$$
\Delta \tau _{\Trm}^{\varphi}(0,d;k)=\Delta \tau _{\Rrm}^{\varphi}(0,d;k)={
\frac{m}{\hbar k\kappa}}\left( {\frac{{2\kappa dk^{2}(\kappa
^{2}-k^{2})+\varepsilon ^{4}\sinh (2\kappa d)}}{{4\kappa
^{2}k^{2}+\varepsilon ^{4}\sinh ^{2}(\kappa d)}}}\right) ,\eqno(3.7)
$$
with $\varepsilon =2mV_{0}/\hbar $. As $d$ increases the term in brackets
approaches 2, and then $\Delta \tau _{\Trm}^{\varphi}$ and $\Delta \tau
_{\Rrm}^{\varphi}$ approach $2m/\hbar k\kappa =2/v_{\grm}\kappa $.
Such a value does not depend at all on the barrier thickness. Therefore,
by increasing the thickness, one can meet
arbitrarily large tunnelling velocities ${d/\tau _{\Trm}^{\varphi}}$: i.e.,
one finds the so-called {\em Hartman--Fletcher effect}[9]
(or simply Hartman effect). \ Let us remark that such an effect
---which has been actually observed in all the previously mentioned
experiments--- implies Superluminal group-velocities for thick ({\em
opaque}) barriers. This seeming disagreement with the so-called Einstein
causality is justified by many authors[10] in terms of
{\em reshaping}$^{\#3}$ of the packet; for a deeper analysis, see ref.[8]
and refs. therein, as well as below.
\footnotetext{$^{\#3}$One has a {\em reshaping} of a packet when,
in crossing the barrier, the lower energy components are, in general,
transmitted less efficiently than those of higher energy. All that can
involve a modification of the form of the packet in the $k$-space with an
effective {\em acceleration}.}

By starting from the phase time, one can study the behaviour
of the various frequencies composing a wavepacket which crosses the barrier.
One purpose is to understand if, and in which cases, reshaping effects
can actually appear, and whether it may be possible to avoid them.
In Fig.3 we represent the values of $T(k)$, computed for different values
of $d$ as a function of $\varepsilon$. This figure shows also the
distribution function of a gaussian packet peaked around $k_0=0.7\varepsilon$,
namely \ $f(k-k_0)={\rm exp}\,[-(k-k_0)^2/2(\Delta k)^2]$, \ with
$\Delta k=0.1\;k_0$. The
weight of each Fourier component in the transmitted packet will be given by
the product $T(k)\;f(k-k_0)$. For very thin barriers, $(d\varepsilon \ll 1)$ $
\;T(k)$ is quasi-constant and very close to 1, except for very small $k$.
Therefore, if $k_0/\varepsilon$ is not too close to $0$, the function
$T(k)\;f(k-k_0)$ will always reach its maximum at $k_0$ and the outgoing packet
will present no distortions. In this case the transmission time,
still computed as a phase time, results longer than the equivalent time, and
no problems arise. On increasing the barrier thickness, however, the lower
energy components of the packet will be transmitted worse and worse. In fact,
for sufficiently large values of $d$, quantity $T(k)$ is very small except
when $k/\varepsilon $ is very close to 1 (in which case it increases,
instead, very quickly). For this reason, the packet is expected by many
authors to be accelerated, since the maximum of the transmitted packet will
result to be situated at $k_0^\prime >k_0$. According to those authors, all
happens as if only a subset of frequencies crossed the barrier: namely,
the higher frequencies, propagating at a velocity larger than $v_\grm$
(also before
reaching the barrier itself!). \ Anyway, by choosing suitable initial
conditions, {\em all these reshaping effects can be avoided (and nevertheless the
Superluminal group-velocities do not disappear)}. Let us see how.

First of all, notice that the barrier cannot have any amplifying effect
on each component of the packet, but it only acts as a filter for some of
them.
Then $T(k)$ is limited and can increase only up to a maximum value equal to 1
(as we can see also from Fig.3). Let us also notice that, for $k<\varepsilon$,
function $T(k)$ is monotonically increasing and, for sufficiently thick barriers,
grows very quickly only for values of $k$ close to $\varepsilon $, where
it shows an ``oblique flexus" going on from concavity to convexity.
We have
$$
\lim_{k\to\varepsilon}T(k)={\frac{1}{\sqrt{1+\displaystyle{\frac{
\varepsilon ^{2}d^{2}}{4}}}}},\eqno(3.8)
$$
and
$$
\lim_{k\to\varepsilon}T^{\prime}(k)={\frac{{2d^{2}\varepsilon
(3+d^{2}\varepsilon ^{2})}}{{3(1+d^{2}\varepsilon ^{2})}^{\frac{3}{2}}}}\eqno
(3.9)
$$
holding also for $k>\varepsilon $. In the vicinity of $\varepsilon $ we may
write
$$
T(k)\simeq {\frac{1}{\sqrt{1+\displaystyle{\frac{\varepsilon ^{2}d^{2}}{4}}}}
}+{\frac{{2d^{2}\varepsilon (3+d^{2}\varepsilon ^{2})}}{{3(1+d^{2}
\varepsilon ^{2})}^{\frac{3}{2}}}}(k-\varepsilon )\eqno(3.10)
$$
and therefore, even close to $k=\varepsilon$, quantity $T(k)$ does not
increase faster than $T^{\prime}(k)\,(k-\varepsilon)$. By contrast, with
regard to $f(k-k_{0})$, we have
\[
f(k-k_{0})={\rm exp}\,[-(k-k_{0})^{2}/2(\Delta k)^{2}].
\]
The argument of the exponential is smaller than 1 for $(k-k_{0})<\Delta
k$ and larger than 1 if $(k-k_{0})>\Delta k$: in the latter case,
$f(k-k_{0})$ decreases exponentially with $[(k-k_{0})/\Delta k]^{2}$.
Except for very thin barriers, for which we have already seen that $
T(k)\simeq 1$, we expect that, if $\varepsilon -k_{0}\sim \Delta k$, a large
part of the tunnelling may also be due to transmission of the higher
energy frequencies. Then we expect that even if the $k$-peak moves
forward, it will always be $k_{0}^{\prime}-k_{0}\sim \Delta k$. On the
contrary, by decreasing the energy of the incident packet, i.e., by taking
$k_{0}$ smaller, the contribute of the higher energies to the tunnelling decreases
considerably, since $f(k-k_{0})$ decreases exponentially with
$[(k-k_{0})/\Delta k]^{2}$: And all that happens still faster for increasing
$|\varepsilon-k_{0}|$. To make possible a reshaping of the packet (in order to
get a forward shifting, or a second peak),
it is then necessary that, in a certain interval, $T(k)$ grows very quickly,
and at a larger extent than the decrease of $f(k-k_{0})$. Namely
$$
{\frac{\drm}{\drm k}}[T(k)f(k)]=T^{\prime}(k)f(k-k_{0})+T(k)f^{\prime
}(k-k_{0})>0,\eqno(3.11)
$$
wich is a constraint very difficult to be studied because of the rather
complicate form
of $T^{\prime}(k)$. But, since $f^{\prime}(k-k_{0})=-\displaystyle{\frac{
(k-k_{0})}{(\Delta k)^{2}}}f(k-k_{0})$, and moreover it is always
$f(k-k_{0})>0$,
condition (3.11) reduces to:
$$
T^{\prime}(k)-{\frac{(k-k_{0})}{(\Delta k)^{2}}}T(k)>0.\eqno(3.12)
$$
One can therefore remark that, as expected, for $k<k_{0}$ equation (3.12) is
always satisfied. For $k>k_{0}$, on the contrary, in order that
$T(k)f(k-k_{0})$ be increasing, one has to require
$$
T^{\prime}(k)>T(k){\frac{(k-k_{0})}{(\Delta k)^{2}}}.\eqno(3.13)
$$
But, as already seen, $T^{\prime}(k)$ is limited, and for sufficiently thick
barriers reaches its maximum at $k=\varepsilon $. Then, as far as quantity
$T(k)$ happens to be small, we can always find some values of $\Delta k$ such that
eq.(3.13) does not hold any more: thus the peak remains at $k_{0}$.  This holds
also for very thick barriers,$^{\#4}$ for which the tunnelling probability
becomes infinitesimal.
\footnotetext{$^{\#4}$For $\kappa d=(\varepsilon ^{2}-k^{2})^{1/2}d\gg 1$,
we have $T(\kappa )=T((\varepsilon ^{2}-k^{2})^{1/2})\sim \displaystyle{
\frac{4k\kappa}{\varepsilon ^{2}}}e^{-\kappa d}$, but in such a case it is
also $T^{\prime}(k)\sim e^{-\kappa d}$.} It is sufficient that: \ 1) $
k_{0}$ be not too close to $\varepsilon $, \ and \ 2) the $k$-distribution
be narrow enough around $k_{0}$.

In Fig.4 the plots of $T(k)\,f(k-k_0)$ are shown for different values of the
barrier thickness and of $k_0$. Only for values
of $k_0/\varepsilon$ very close to 1 ($ k_0/\varepsilon \simeq 0.9$), the
tunnel occurs mostly for the above-barrier energy components.

Let us finally notice that, even though the peak should move forward in
the $k-$space, this would not have any direct influence on its position in
the $x$-space; actually, in that case, either $\tau_{\Trm ,\Rrm}^\varphi$ or
$\Delta\tau_{\Trm,\Rrm}^\varphi$ would lose any physical sense, since
they have been evaluated for $k=k_0$. \ This ensures that {\em the transmitted
packet does not suffer sensible distortions with respect to the incident
one}.

\

\section{Times built-up following the centroid method} 

Some authors, rather than following the peak through the stationary phase method,
prefer to refer to {\em the centroid of the packet}. This because, when applying the
stationary phase method, we are forced to employ packets very narrow in $k$
(and then very extended in $x$). And also because, in so doing, we can
better evaluate the effects of the possible acceleration caused by the
barrier crossing.

Suppose the packet be initially $(t\leq 0)$ located at a certain
distance from the barrier, so that \ $\int_{0}^{\infty}|\psi (x,0)|^{2}\drm
x\simeq 0$, \ that is, the probability is negligible that the particle lies
on the right of 0. We identify the ``particle position" at time $t=0$ with
the position of its {\em center-of-mass} (the average space-coordinate):
$$
\bar{x}(0)=\displaystyle{\frac{{\int_{-\infty}^{\infty}x|\psi (x,0)|^{2}dx}
}{{\int_{-\infty}^{\infty}|\psi (x,0)|^{2}dx}}}\,.\eqno(4.1)
$$
It being
$$
f(k)=f(k,0)\displaystyle{\frac{1}{\sqrt{2\pi}}}\int dx\psi
(x,0)e^{-ikx}=|f(k)|e^{i\xi (k)} \ , \eqno(4.2)
$$
it is possible to show that[11]
$$
x_{0}=\bar{x}(0)={\frac{{-\int_{0}^{\infty}\drm k|f(k)|^{2}\displaystyle{
\frac{{\drm\xi}}{{\drm k}}}}}{{\int_{0}^{\infty}\drm k|f(k)|^{2}}}}=-\lan\xi
^{\prime}(k)\ran\,.\eqno(4.3)
$$
We'll have
$$
\bar{x}_{\Inrm}(t)=x_{0}+{\frac{\hbar}{m}}\lan k\ran_{\Inrm}t,\qquad (t\to 0);\eqno(4.4)
$$
$$
\bar{x}_{\Trm}(t)=x_{0}+{\frac{\hbar}{m}}\lan k\ran_{\Trm}t-
\lan\alpha^{\prime}\ran_{
\Trm},\qquad (t\to\infty );\eqno(4.5)
$$
$$
\bar{x}_{\Rrm}(t)=x_{0}+{\frac{\hbar}{m}}\lan k\ran_{\Rrm}t-
\lan\beta^\prime\ran_\Rrm \qquad (t\to\infty );\eqno(4.6)\,,
$$
where
\[
\lan k\ran_{\Inrm}\equiv \frac{\int_{-\infty}^{\infty}\drm k\,k|f(k)|^{2}}{
\int_{-\infty}^{\infty}\drm k|f(k)|^{2}}\qquad \quad \lan k\ran_{\Trm}\equiv
\frac{\int_{-\infty}^{\infty}\drm k\,k|T|^{2}}{\int_{-\infty}^{\infty}
\drm k|T|^{2}}\qquad \quad \lan k\ran_{\Rrm}\equiv \frac{\int_{-\infty}^{\infty}
\drm k\,k|R|^{2}}{\int_{-\infty}^{\infty}\drm k|R|^{2}}
\]
\[
\lan\alpha^\prime\ran_{\Trm}\equiv \frac{\int_{-\infty}^{\infty}\drm
k\lan\alpha^\prime\ran T|^{2}}{\int_{-\infty}^{\infty}\drm k|T|^{2}}\qquad
\quad \lan\beta ^{\prime}\ran_{\Rrm}\equiv \frac{\int_{-\infty}^{\infty}\drm
k\lan\beta^\prime\ran |R|^{2}}{\int_{-\infty}^{\infty}\drm k|R|^{2}}\,.
\]
On shifting $\bar{x}_{\Inrm}(t)$ and $\bar{x}_{\Rrm}(0)$ forward in time,
and $\bar{x}_{\Trm}(t)$ backwards, it is possible to extrapolate $t_{\Inrm
}(0)$, $t_{\Rrm}(0)$ and $t_{\Trm}(d)$, respectively, to times at which the
centroid passes$^{\#5}$ through 0 and $d$.
\footnotetext{$^{\#5}$Notice that we are still using asymptotic forms of the
wave-function, always neglecting, then, the self-interference effects near the
barrier.} Then the transmission and reflection times will be given by
$$
\tau _{\Trm}^{\Crm}=t_\Trm(d)-t_\Inrm(0)=\frac{m}{\hbar}\,
\left[\frac{d-x_{0}+\lan\alpha^\prime\ran_\Trm}{\lan k\ran_\Trm} +
\frac{x_{0}}{\lan k\ran_\Inrm}\right]\,;\eqno(4.7)
$$
$$
\tau _{\Rrm}^{\Crm}=t_{\Rrm}(0)-t_{\Inrm}(0)={\frac{m}{\hbar}}\left[ {\frac{
{-x_{0}+\lan\beta^\prime\ran_{\Rrm}}}{{\lan k\ran_\Rrm}}}+
{\frac{{x_{0}}}{{\lan k\ran_{ \Inrm}}}}\right]\,.\eqno(4.8)
$$
Leavens and Aers[12] showed that, for $\Delta k\to 0$, we have
$\tau_{\Trm}^{\Crm}\to\Delta\tau_{\Trm}^{\varphi}$ and
\mbox{$\tau_{\Rrm}^{\Crm}\to\Delta\tau_{\Rrm}^{\varphi}$}, while
the possible corrections are of the first order in $\Delta k$, according
to our previous qualitative reasonings.

Analogous results where achieved by Martin and Landauer[13], who
performed the calculations in the electromagnetic framework, and by Collins,
Lowe and Barker[4] who followed the time evolution of a gaussian packet
obeying the time-dependent Schr\"{o}dinger equation. Nevertheless, they also
got, even by explicit computations, the time spent by the centroid to leave
the barrier (there are no self-interferences in the wave-function
for $x\geq d$), by {\em extrapolating} the time spent to reach the
barrier.[6]

Let us go on now to study some tunnelling times defined by means of
suitable ``clocks".

\

\section{B\"{u}ttiker and Landauer times} 

In order to determine $\tau_\Trm$, B\"{u}ttiker and Landauer[14-16] in
1982 proposed to consider a time-oscillating square-barrier, and supposed
the crossing time to equal the duration of the interaction between the particle
and the oscillating potential.

Let us consider a square potential with height $V_0$, upon which an
oscillating potential $\delta V\cos\omega t$ is superimposed. At very low
frequencies, the potential varies very slowly; therefore the particle,
during the crossing, will feel the effects of only a part of the modulation
cycle. So, as long as the period corresponding to the oscillation is long
compared to the crossing time, the particle will interact with a quasi-static
potential. At higher frequencies $(\omega\gg 1/\tau_\Trm)$, the particle
will undergo the effect of several cycles of oscillation and will absorb or
release energy quanta equal to $\hbar\omega$. The transition frequency between
the adiabatic behaviour, typical of the low frequencies, and the non-adiabatic
behaviour, yields an approximate measure of the duration of the particle
interaction with the barrier.

At the first order in $\delta V$, only the two bands $E{\pm} \hbar\omega$
will appear. Moreover, the particles belonging to the higher energy band will
have a larger crossing-probability, with respect to the low energy ones.
B\"{u}ttiker and Landauer showed that, for thick barriers and not too high
frequencies ($\hbar\omega$ small compared with either $E$ or $V_0-E$), the
relative intensity of the two bands will be given by:
$$
I_{\pm}^\Trm(\omega)=\left|{\frac{{T_{\pm}(\omega)}}{{T(\omega)}}}\right|^2
=\left({\frac{{\delta V\tau_\Trm^{\BLrm}}}{{2\hbar\omega}}}\right)^2 \left[
e^{{\pm}\omega {\frac{md}{\hbar \kappa}}}-1\right]^2. \eqno(5.1)
$$
Then those two authors identified the crossing time with $\tau_\Trm^{\BLrm}=
\displaystyle{md / (\hbar \kappa)}$. In the very small frequency limit,
eq.(5.1) reduces to
$$
\left|{\frac{{T_{\pm} (\omega )}}{{T(\omega )}}}\right|^2  =\left({\frac{{
\delta V\tau_\Trm^{\BLrm}}}{{2\hbar}}}\right)^2\,. \eqno(5.2)
$$
As expected, the number of particles that will have absorbed or released
energy is, in this case, quite independent of $\omega$. Again, from eq.(5.1),
we have:
$$
{\frac{T_+(\omega)-T_-(\omega)}{T_+(\omega)+T_-(\omega)}}=\tanh
(\omega\tau_\Trm^{\BLrm})\,.\eqno(5.3)
$$
This shows that just $\tau_\Trm^{\BLrm}$ determines the transition from the
adiabatic behaviour at low frequencies, when $T_+\simeq T_-$, to the high
frequencies behaviour, when $T_+\gg T_-$. With regard to the reflected
particles, always within the limits $\hbar\omega\ll E$ \ and \
$\mbox{$\quad\hbar\omega\ll V_0-E$}$, B\"{u}ttiker and Landauer found that
$$
\left|{\frac{R_{\pm}}{R}}\right|^2=\left({\frac{\delta V\tau_\Rrm^{\BLrm}}{
2\hbar}}\right)^2\,, \eqno(5.4)
$$
with $\tau_\Rrm=\displaystyle{\hbar k / (V_0\kappa)}$. Notice that also
eq.(5.4) is independent of $\omega$.

The low frequencies behaviour of eqs.\,(5.2) and (5.4) is typical of a system
characterized by two-states, $|1\ran$ and $|2\ran$, endowed with energy $E$
and $E{\pm}\hbar\omega$ respectively, brought to resonance by a perturbation
$V_1\cos\omega t$. If for $t=0$ the whole population of the system is in the
state $|1\ran$, the population of the state $|2\ran$ increases initially as
\ $(V_1 t/2\hbar)^2$. The same happens if the energies of the two levels
$E_1$ and $E_2$ are equal. Thus, in eqs.(5.2) and (5.4), quantities
$\tau_\Trm^{\BLrm}$ and $\tau_\Rrm^{\BLrm}$ actually play the role of
interaction times of the particle-barrier system.

\

\section{Larmor times} 

In 1966 Baz'\,[17,18] proposed to exploit the Larmor precession, caused by
the presence of a magnetic field on particles {\em endowed with spin}, to
measure
the collision times of these particles. In the same year, Rybachenko[19]
applied this method to compute the tunnelling times for a
one-dimensional square barrier. Let us consider a beam of spin-${\frac{1}{2}}$
particles, polarized along $\hat{x}$, with mass $m$ and kinetic energy $E$,
moving along $\hat{y}$ (see Fig.5; notice that, following the
original paper, also in this figure the propagation axis has been named $y$). \
Let us furthermore suppose that a weak, homogeneous magnetic field, $\Bbf_{0}$, directed along the
$\hat{z}$-axis, is present in the barrier zone, overlapping the barrier potential.
Following Rybachenko, the particles which enter the barrier, on crossing
the magnetic field will undergo a Larmor precession with frequency $\omega
_{\Lrm}=g\displaystyle{\mu B_{0}/\hbar}$, where $g$ is the gyromagnetic
ratio, and $\mu $ the magnetic moment. The precession will stop just when
the particle will come out from one of the two sides of the barrier.
Since we have[20]
\[
\lan S_{x}\ran_{\Trm}={\frac{\hbar}{2}}\cos \omega _{\Lrm}\tau ^{\Lrm},
\]
\[
\lan S_{y}\ran_{\Trm}=-{\frac{\hbar}{2}}\sin \omega _{\Lrm}\tau ^{\Lrm}\,,
\]
we can write, in the weak magnetic field limit,
\[
\lan S_{x}\ran_{\Trm}\simeq {\frac{\hbar}{2}},
\]
\[
\lan S_{y}\ran_{\Trm}\simeq -{\frac{\hbar}{2}}\omega _{\Lrm}\tau _{y\Trm}^{\Lrm}\,.
\]
Thus the nonzero spin component along $\hat{y}$ will be proportional to the
dwell time inside the barrier. As a consequence:
\[
\tau _{y\Trm}^{\Lrm}=\lim_{\omega _{\Lrm}\to 0}{\frac{\lan S_{y}\ran_{\Trm}
}{{-{\frac{1}{2}}\hbar \omega _{\Lrm}}}}\,.
\]
Rather strangely, Rybachenko did not consider the main effect of the field
on the particles, namely the spin alignment. As a matter of fact, after having
left the barrier, the spin will have a component along $\hat{z}$
equal to ${\pm}\hbar/2$.  While outside the barrier the particle energy
is independent of the spin, on the contrary inside the barrier its energy
will depend also on the spin $z$-component (because of the Zeeman effect).
The difference
between the spin-up and spin-down energies is ${\pm}\hbar \omega_{\Lrm}/2$.
Let us state:
\[
\psi _{\inrm}={\frac{1}{\sqrt{2}}}\left(
\matrix{1\cr 1}\right) e^{iky}
\]
\[
\psi _{\Trm}=(|D_{+}|^{2}+|D_{-}|^{2})^{-1/2}\left(
\matrix{D_+\cr D_-}\right) e^{iky}
\]
with
\[
D_{\pm}=T(\kappa _{\pm})e^{\alpha}e^{-i\kappa _{\pm}d}\,,
\]
where $\kappa _{\pm}$ are the $\kappa$-values corresponding to $E{\pm}
\displaystyle{\frac{\hbar}{2}}\omega _{\Lrm}$. \
If
\[
\lan S_{\irm}\ran_{\Trm}={\frac{\hbar}{2}}\lan\psi|\hat{\sigma}_{\irm}|\psi\ran\,,
\]
we obtain
$$
\lan S_{z}\ran_{\Trm}={\frac{\hbar}{2}}{\frac{|T_{+}|^{2}-|T_{-}|^{2}}{
|T_{+}|^{2}+|T_{-}|^{2}}},\eqno(6.1a)
$$

$$
\lan S_y\ran_\Trm=-\hbar\sin (\alpha_+-\alpha_-){\frac{|T_+ T_-|}{|T_+|^2+|T_-|^2}}
, \eqno(6.1b)
$$

$$
\lan S_x\ran_\Trm=\hbar\cos (\alpha_+-\alpha_-){\frac{|T_+ T_-|}{|T_+|^2+|T_-|^2}}
\,.\eqno(6.1c)
$$
Analogous expressions are found for the reflected particles: it is sufficient
to replace $T_{\pm}$ by $R_{\pm}$. Furthermore, it can be shown that
$$
\lan S_z\ran_\Rrm=-\lan S_z\ran_\Trm{|T_+|^2 -|T_-|^2\over |R_+|^2+|R_-|^2}, \eqno(6.2a)
$$

$$
\lan S_y\ran_\Rrm=-\lan S_y\ran_\Trm\left|{\frac{R_+R_-}{T_+T_-}}\right| {\frac{|T_+|^2
+|T_-|^2}{|R_+|^2+|R_-|^2}}\eqno(6.2b)
$$

$$
\lan S_x\ran_\Rrm=-\lan S_x\ran_\Trm\left|{\frac{R_+R_-}{T_+T_-}}\right| {\frac{|T_+|^2
+|T_-|^2}{|R_+|^2+|R_-|^2}}\,.\eqno(6.2c)
$$
For a weak magnetic field we have \ $\kappa_{\pm}\simeq \kappa\mp
\displaystyle{m\omega_\Lrm / \hbar},$ \ and also
$$
|T_+|^2 -|T_-|^2\sim -{\frac{m\,\omega_\Lrm}{\hbar\kappa}} {\frac{\partial T}{
\partial\kappa}}\,. \eqno(6.3)
$$
In (6.3) the term \ $-m(\hbar\kappa)^{-1}\,(\partial T/\partial\kappa)$, \
which multiplies $\omega_\Lrm$, has, of course, the dimensions of a time.
In 1983 B\"{u}ttiker suggested[21] the introduction of three times, namely
$\tau_{z\Trm}^\Lrm$, $\tau_{y\Trm}^\Lrm$ and $\tau_{x\Trm}^\Lrm$, in the
following way. He assumed
$$
\lan S_z\ran_\Trm=(\hbar/2)\omega_\Lrm\tau_{z\Trm}^\Lrm, \eqno(6.4a)
$$

$$
\lan S_y\ran_\Trm=-(\hbar/2)\omega_\Lrm\tau_{y\Trm}^\Lrm, \eqno(6.4b)
$$

$$
\lan S_x\ran_\Trm=(\hbar/2)[1-(\omega_\Lrm^2{\tau_{x\Trm}^\Lrm}^2)/2], \eqno(6.4c)
$$
and then:
$$
\tau_{z\Trm}^\Lrm=\lim_{\omega_\Lrm\to 0} {\frac{\lan S_z\ran_\Trm}{{{\frac{\hbar}{2
}}\omega_\Lrm}}}=  -{\frac{m}{\hbar\kappa}}{\frac{\partial \ln T}{
\partial\kappa}}, \eqno(6.5a)
$$

$$
\tau_{y\Trm}^\Lrm=\lim_{\omega_\Lrm\to 0} {\frac{\lan S_z\ran_\Trm}{{{\frac{\hbar}{2
}}\omega_\Lrm}}}=  -{\frac{m}{\hbar\kappa}}{\frac{\partial\alpha}{
\partial\kappa}}, \eqno(6.5b)
$$

$$
\tau_{x\Trm}^\Lrm=\lim_{\omega_\Lrm\to 0} {\frac{\lan S_z\ran_\Trm}{{{\frac{\hbar}{2
}}\omega_\Lrm}}}= {\frac{m}{\hbar\kappa}}\left[ \left({\frac{\partial\alpha}{
\partial\kappa}}\right)^2 + \left({\frac{\partial \ln T}{\partial\kappa}}
\right)^2\right]. \eqno(6.5c)
$$
On developing the calculations, B\"{u}ttiker obtained
$$
\tau_{z\Trm}^\Lrm={\frac{m\varepsilon^2}{\hbar\kappa^2}} {\frac{
(\kappa^2-k^2)\sinh^2(\kappa d)+  (\kappa d\varepsilon^2/2)\sinh(2\kappa d)
}{ 4k^2\kappa^2+\varepsilon^4\sinh^2(\kappa d)}} \eqno(6.6a)
$$
$$
\tau_{y\Trm}^\Lrm={\frac{mk}{\hbar\kappa^2}} {\frac{2\kappa d(\kappa^2-k^2)
+ \varepsilon^2\sinh (2\kappa d)}{ 4k^2\kappa^2+\varepsilon^4\sinh^2(\kappa
d)}} \eqno(6.6b)
$$
$$
\tau_{x\Trm}^\Lrm=\sqrt{\tau_{z\Trm}^2+\tau_{y\Trm}^2} \eqno(6.6c)
$$
For sufficiently thick barriers he got
$$
\tau_{z\Trm}^\Lrm\simeq {\frac{md}{\hbar\kappa}},\qquad\quad  \tau_{y\Trm
}^\Lrm\simeq {\frac{2mk}{\hbar\varepsilon^2\kappa}}\,. \eqno(6.7)
$$
In such a way, in the thick barriers limit, $\tau_{z\Trm
}^\Lrm=\tau_\Trm^{\BLrm}$\,. After all, we are not confronting a real spin
precession, but just a spin ``flip'' together with a splitting of the energy
levels. B\"{u}ttiker himself[21] showed that for $\tau_{z\Trm}^\Lrm$
there hold considerations analogous to the ones holding for $\tau_\Trm^{
\BLrm}$. In connection with $\tau_{x\Trm}^\Lrm$, it is even more
difficult to recognize a physical meaning in it. Indeed, if we think that
also the spin $x$-component precedes around the $\hat z$-axis, therefore to
such a precession it should correspond an average spin $x$-component equal to
\[
\lan S_x\ran_\Trm=(\hbar/2)[1-(\omega_\Lrm^2{\tau_{y\Trm}^\Lrm}^2)/2],
\]
and {\em not} to
$$
\lan S_x\ran_\Trm=(\hbar/2)[1-(\omega_\Lrm^2{\tau_{x\Trm}^\Lrm}^2)/2]\,.\eqno(6.4c)
$$
For the previous considerations about $\tau_{z\Trm}^\Lrm$, quantity $\tau_{x
\Trm}^\Lrm$ could be regarded, at most, as an average of $\tau_{y\Trm}^\Lrm$
and $\tau_{z\Trm}^\Lrm$; \ and, in fact, some authors introduce directly a time
\ $\tau_\Trm^\Brm=\displaystyle\sqrt{{\tau_{y\Trm}^\Lrm}^2 + {\tau_{z\Trm
}^\Lrm}^2}$.\ In the thick barriers limit, we have $\tau_\Trm^\Brm\simeq
\tau_{z\Trm}^\Lrm$. Therefore, the only one of the three {\em Larmor times}
which seems endowed with a clear physical meaning is
$\tau_{y\Trm}^\Lrm$. Associated with this quantity, Falk and Hauge[22] found,
in 1988, the two relations:
$$
\tau_{y\Trm}^\Lrm={\frac{m}{\hbar k}}(x_2-x_1+\alpha^\prime)+{\frac{mR}{
2\hbar k^2}}  [\sin (\beta -2kx_1)-\sin (2\alpha - \beta + 2kx_2)] \ , \eqno
(6.8a)
$$
as well as, for the reflected part,
\[
\tau_{y\Rrm}^\Lrm={\frac{m}{\hbar k}}(x_2-x_1+\alpha^\prime)+{\frac{mR}{
2\hbar k^2}}\,[\sin (\beta -2kx_1)-\sin (2\alpha - \beta + 2kx_2)] +
\]
$$
+ {\frac{m}{2\hbar k^2R}}  [\sin (\beta -2kx_1)+\sin (2\alpha - \beta +
2kx_2)] \ , \eqno(6.8b)
$$
where $x_1$ and $x_2$ are any pair of points outside the
barrier (one on its right and one on its left), and the magnetic field,
rather than limited to the mere barrier zone, covers the whole range
($x_1$,$x_2$). It can be easily seen that eqs.(6.7) are related
to the phase time much more than the oscillating terms, whose amplitudes
increase when the incident energy decreases. For a square barrier we obtain
\[
\tau_{y\Trm}^\Lrm(d)=\tau_{y\Rrm}^\Lrm(d)=\tau^\Drm(d),
\]
where $\tau^\Drm$ is nothing but the {\em dwell time}, which will be
introduced in the following.

\

\section{Complex times: path-integrals} 

The introduction of complex times follows initially from the idea that for
above-barrier energies we can write \ $v=\hbar\kappa=\hbar\displaystyle\sqrt{
k^2-\varepsilon^2}$ \ and, therefore, $\tau_\Trm=d/v=md/\hbar\kappa$. For
$E<V_0$, on the contrary, the wave-vector becomes imaginary. Nevertheless,
let us imagine the below-barrier motion of a particle to occur along
a classical trajectory, but with {\em imaginary velocity} and {\em time}. We
can realize that the trajectories have to be complex, whilst times and velocities
may be assumed as real, by considering the quantities:
\[
v^\Srm={\frac{\hbar |\kappa |}{m}},
\]
\[
\tau_\Trm^\Srm={\frac{d}{v}}={\frac{md}{\hbar \kappa}},
\]
where label {\rm S} means ``semiclassical''. Of course, also $\tau_\Trm^\Srm$ is
not physically meaningful, since it diverges for $k=\varepsilon$. Anyhow, notice
that in the case of thick barriers also $\tau_\Trm^{\BLrm}$, and of
course $\tau_{z\Trm}^\Lrm$, approach $\tau_\Trm^\Srm$. Let us recall that in
1992 Hagmann[23] proposed to consider the case of a particle that, in order to
cross the barrier, receives a certain energy $\Delta E$, during a time
interval $\Delta t$. Even if that procedure seems to lack physical meaning, his
result (obtained by applying the uncertainty principle) is correct: \ $\Delta t=\tau_\Trm^\Srm =
md/\hbar\kappa$\,.

Going back to the complex times, in 1987 Sokolovski and Baskin[24] set
forward a generalization of the classical concept of time to quantum
mechanics and hence applied their method to the tunnel effect. Let us
consider a particle that, emitted at the point $\rbf_1$ at time $t_1$,
is detected at the point $\rbf_2$ at time $t_2$.  Moreover, suppose
that the particle, moving along the trajectory $\rbf(t)$ inside a potential
$V(\rbf)$, has crossed a certain space region $\Omega$. Then the time spent by
the particle in that space region will be given by
$$
\tau_{\clrm}^\Omega =\int_{t_1}^{t_2} \drm t\,\Theta_\Omega (\rbf(t)),\eqno
(7.1)
$$
where quantity $\Theta_\Omega (\rbf(t))$ is 1 if $\rbf(t)$ belongs to $\Omega
$, and 0 otherwise. In the one-dimensional case, we'll have
$$
\tau_{\clrm}^\Omega =\int_{t_1}^{t_2} \drm t \int_0^d dx\ \delta (x-x(t)).
\eqno(7.2)
$$
If we then use the Feynman path-integral method to build-up trajectories
along which to perform the time averages, we get
\[
\tau^\Omega (x_1,t_1;x_2,t_2;k)=\lan\tau_{\clrm}^\Omega[x({\cdot})]\ran_{{\rm path
}},
\]
where $x({\cdot})$ is an arbitrary path (in the phase-space) between $
(x_1,t_1)$ and $(x_2,t_2)$. In general $\tau^\Omega $ will be complex.
Sokolovski and Baskin found
$$
\tau_\Trm^\Omega =i\hbar\int_0^d dx\ {\frac{\delta \ln A}{\delta V(x)}},\eqno
(7.3a)
$$

$$
\tau_\Rrm^\Omega =i\hbar\int_0^d dx\ {\frac{\delta \ln B}{\delta V(x)}},\eqno
(7.3b)
$$
with $A=T e^{i\alpha}$,\ \ $B= R e^{i\beta}$. In the same paper, those authors
wrote between $\tau^\Omega$ and the Larmor times the relations 
$$
\Rerm \ \tau_\Trm^\Omega =\tau_{y\Trm}^\Lrm,\eqno(7.4a)
$$
$$
\Imrm \ \tau_\Trm^\Omega =\tau_{z\Trm}^\Lrm,\eqno(7.4b)
$$
$$
|\tau_\Trm^\Omega| =-\tau_{x\Trm}^\Lrm.\eqno(7.4c)
$$
Analogously, for the reflected part:
$$
\Rerm \ \tau_\Rrm^\Omega =\tau_{y\Rrm}^\Lrm,\eqno(7.5a)
$$
$$
\Imrm \ \tau_\Rrm^\Omega =\tau_{z\Rrm}^\Lrm,\eqno(7.5b)
$$
$$
|\tau_\Rrm^\Omega| =\tau_{x\Rrm}^\Lrm.\eqno(7.5c)
$$
In spite of such a surprising correlation between $\tau^\Omega$ and the Larmor
times, it is difficult to physically interpret such results. A possible
interpretation$^{\#6}$ is the one given by H\"{a}nggi[25] in 1993.
\footnotetext{$^{\#6}$Apparently meaningless, but probably not that much as it
seems; recall, e.\,g., the Caldirola chronon: cf. R.H.A.Farias and E.Recami,
``Introduction of a quantum of Time (`chronon') and its consequences for
quantum mechanics", LANL Arhives e-print \# quant-ph/9706059.} According to
that author, the tunnelling time would be characterized by two time scales.
However, he found it difficult to justify his approach from the physical
point of view. Sokolovski and Connor[26], the same year,
criticized that theory: they regarded the existence of two
different crossing times as physically doubtful; and concluded that what
is to be considered as crossing time is the magnitude of $\tau_\Trm^\Omega $.

Let us go back to H\"{a}nggi's hypothesis. If we look at the transmitted
wave-form, we can point out that
$$
\psi_\Trm=\psi_{\IIIrm}(x,k)=T(k)e^{i\alpha}e^{ikx}=e^{\ln
T(k)}e^{i\alpha}e^{ikx}\,. \eqno(7.6)
$$
Notice that, when applying the phase time definition to the above equation,
one finds two distinct components: one proportional to $\drm\alpha/\drm E$, and
one to $\drm(\ln T)/\drm E$. We can then imagine the latter to be the time
needed to damp the signal during the crossing. It appears strange the dependence of
such a time on $d$, since $\tau_\Trm^\Omega$, rather than decreasing when the
barrier thickness increases, does increase proportionally to it. It should be 
better understood how such a time may be linked to the level transitions, both
in the B\"{u}ttiker-Landauer theory, and in the spin-flip case.

\

\section{Complex times: Bohm's method} 

Still in the complex-times approach, Leavens and Aers[27] in 1993, starting
from the same operator introduced by Sokolowski and Baskin ($\tau_{\clrm
}^{\Omega}$), suggested to have recourse to {\em Bohm mechanics}.
Bohm mechanics yields ``semiclassical'' trajectories, which can be employed to
compute the average tunnelling time. The Bohm method provides us with a
couple of equations fully equivalent to the Schr\"{o}dinger equation,
simultaneously allowing a rather "classical" interpretation of quantum theory.
Let us resume it shortly, following previous work of ours.

The most general scalar wave-function $\psi \in \,$I$\!\!\!$C may be
factorized as follows:
$$
\psi =\sqrt{\rho}\,\,{\rm exp}\!\left(i\,\frac{\varphi}{\hbar}\right)\,,\eqno(8.1)
$$
where $\rho(\xbf,t),\,\varphi(\xbf,t)\in$I$\!$R. \ Applying this assumption in
the Schr\"odinger equation, and separating the real from the imaginary part,
we easily get the two well-known equations[28] for the so-called Madelung
probabilistic fluid (which, taken together, are equivalent to the
Schr\"{o}dinger equation), i.\,e.:
$$
\pa_{t}\varphi +\frac{1}{2m}({\vecna}\varphi )^{2}+\frac{\hbar ^{2}}{4m}
\left[ \frac{1}{2}\left( \frac{{\vecna}\rho}{\rho}\right) ^{2}-\frac{
\triangle \rho}{\rho}\right] +U=0,\eqno(8.2)
$$
where
$$
\frac{\hbar ^{2}}{4m}\left[ \frac{1}{2}\left( \frac{{\vecna}\rho}{\rho}
\right) ^{2}-\frac{\triangle \rho}{\rho}\right] \equiv -\frac{\hbar ^{2}}{
2m}\frac{\triangle |\psi |}{|\psi |}\eqno(8.3)
$$
is often called ``quantum potential''; and
$$
\pa_{t}\rho +{\vecna}{\cdot}(\rho {\vecna}\varphi /m)=0.\eqno(8.4)
$$
Equations (8.2), (8.4) are the {\em Hamilton--Jacobi} and the {\em continuity}
equations for the non-relativistic probabilistic fluid, respectively, and
constitute the ``hydrodynamic'' formulation of the Schr\"{o}dinger theory. Once
chosen the initial and the boundary conditions, the solution of the Madelung
equation system yields the semiclassical phase $\varphi $ and the probability
density $\rho$. Matching the quantum phase with the classical action, and
neglecting the presence of spin, Bohm assumed for the particle ``local''
velocity the expression
$$
\vbf(\xbf,t)=\frac{\imp(\xbf,t)}{m}=\frac{\vecna\varphi(\xbf,t)}{m}\,.\eqno(8.5)
$$
Therefore, by integrating the above velocity field, we can compute the
semiclassical Bohm trajectories, and from them derive {\em semiclassical
tunnelling times}.

\

\section{Dwell time} 

The {\em dwell time} was introduced first by Smith[29] in 1960, in order
to estimate the average duration of a collision process without
distinguishing among the various channels. As already said, it is defined as
the ratio between the probability that the particle {\em is} in a certain
region of space and the flux $j_{\inrm}$ entering that same region, without
taking into account if the particle is reflected or transmitted:
$$
\tau^\Drm(x_1,x_2;k)=j_{\inrm}^{-1}\int_{x_1}^{x_2}|\psi (x,k)|^2 dx  = {
\frac{1}{v_\grm}}\int_{x_1}^{x_2}|\psi (x,k)|^2 dx. \eqno(1.1)
$$
For a square barrier, we have
$$
\tau^\Drm(x_1,x_2;k)={\frac{mk}{\hbar\kappa^2}} {\frac{2\kappa
d(\kappa^2-k^2) + \varepsilon^2\sinh (2\kappa d)}{ 4k^2\kappa^2+
\varepsilon^4\sinh^2(\kappa d)}} \ , \eqno(9.1)
$$
which, for $\kappa d\gg 1$, becomes
$$
\tau^\Drm= {\frac{\hbar k}{V_0\kappa}}={\frac{2mk}{h\varepsilon^2\kappa}} \ .
\eqno(9.2)
$$
Thus, for sufficiently thick barriers, also $\tau^\Drm$, like $\tau_{\Trm,\Rrm
}^\varphi$ and $\Delta\tau_{\Trm,\Rrm}^\varphi$, turns out to be {\em
independent of the thickness}, while it decreases for increasing $k$, and vanishes for $k=0$.
Assuming that transmission and reflection are situations excluding each other,
almost all the authors concluded that any dwell time must necessarily
satisfy the relation
$$
\tau _{\Drm}=|R(k)|^{2}\tau_{\Rrm} + |T(k)|^{2}\tau_{\Trm}\,.\eqno(1.2)
$$
Such a relation is, for instance, satisfied by $\tau ^{\Omega}$. Indeed
Sokolovski and Baskin[24], in 1987, found that
\[
\tau ^{\Drm}=|R|^{2}\tau _{\Rrm}^{\Omega}+|T|^{2}\tau _{\Trm}^{\Omega}.
\]
From the above equation, by separating the real from the imaginary
part, we get
\[
\cases{\tau^\Drm=|R|^2\tau_{y\Rrm}^\Lrm+|T|^2\tau_{y\Trm}^\Lrm\cr \ \cr
         |R|^2\tau_{z\Rrm}^\Lrm+|T|^2\tau_{z\Trm}^\Lrm=0.\cr}
\]
The former equation was obtained independently by Falck and Hauge[22] one year
later. The latter equation, instead, is nothing but the conservation law
of an angular momentum: in the present case, of the spin $z$-component.
Equation (9.1) does not appear to be satisfied for the phase time and, 
a fortiori, for the extrapolated phase times. Actually, Hauge et al.[11] found
the relation
$$
\tau ^{\Drm}(x_{1},x_{2};k) =
|T(k)|^{2}\tau _{\Trm}^{\varphi}(x_{1},x_{2};k)+|R(k)|^{2}\tau _{\Rrm
}^{\varphi}(x_{1},x_{2};k)+{\frac{mR}{\hbar k^{2}}}\sin (\beta -2kx_{1})\,.
\eqno(9.3)
$$
Yet, those two authors showed also that, if one considers that any packet has a
certain spread in $k$ and applies eq.(9.3) to the whole packet, one gets$^{\#7}$
\footnotetext{$^{\#7}$Of course, in eq.(9.4) we should consider
$\lan|T(k)|^{2}\tau_{\Trm}^{\varphi}(x_{1},x_{2};k)\ran$ and
$\lan|R(k)|^{2}\tau_{\Rrm}^{\varphi}(x_{1},x_{2};k)\ran$, rather than
$\lan|T(k)|^{2}\ran\lan\tau_{\Trm}^{\varphi}(x_{1},x_{2};k)\ran$ and
$\lan|R(k)|^{2}\ran\lan\tau_{\Rrm}^{\varphi}(x_{1},x_{2};k)\ran$. But it can
be proved[4] that the error made by using eq.(9.5) is simply of the same
order as $\sigma$.}
$$
\lan\tau ^{\Drm}(x_{1},x_{2};k)\ran\simeq
\lan|T(k)|^{2}\ran\lan\tau_{\Trm}^{\varphi}(x_{1},x_{2};k)\ran +
\lan|R(k)|^{2}\ran\lan\tau_{\Rrm}^{\varphi}(x_{1},x_{2};k)\ran +
$$
$$
+\,{\frac{mR}{\hbar k^{2}}} \; \sigma ^{-1}\int \drm k\sin (\beta
-2kx_{1})+O(\sigma)\,.\eqno(9.4)
$$
Thus, if $|x_{1}|\gg \sigma ^{-1}$, the argument of the integral will
oscillate quickly enough to make it negligible, and then we can write
$$
\lan\tau ^{\Drm}(x_{1},x_{2};k)\ran \,=\,
\lan|T(k)|^{2}\ran\lan\tau_{\Trm}^{\varphi}(x_{1},x_{2};k)\ran +
\lan|R(k)|^{2}\ran\lan\tau_{\Rrm}^{\varphi}(x_{1},x_{2};k)\ran.\eqno(9.5)
$$
This is an equation agreeing with eq.(1.2) up to ${\rm O}(\sigma )$. Actually
eq.(9.5) does not hold for the extrapolated phase times, in such a way
showing once more the purely asymptotic nature of the phase time.

Let us analyze a little better eq.(9.3). According to Hauge and St\o vneng[4], this
equation shows that $\tau^\Drm$ represents just the exact time spent by the
particles inside the barrier, while the term $mR{(\hbar k^2)}^{-1}\sin(\beta
-2kx_1)$ represents a $\Delta\tau$ caused by self-interference effects.
In fact, the dwell time computed in the interval $(-L,x_1)$ does diverge
when $L$ grows to infinity.
When subtracting the dwell time in $(-L,x_1)$ computed only for the {\em incident}
part of the wavepacket, we obtain for$^{\#8}$ $\Delta\tau^\Drm(x<x_1;k)$, after
some algebra: \footnotetext{$^{\#8}$A similar approach, in which positive
and negative fluxes are however evaluated separately, was later adopted
by Olkhovsky and Recami in order to generalize the definition of dwell time:
cf. refs.[5], and [8].}
$$
\Delta\tau^\Drm(x<x_1;k)=-{\frac{mR}{\hbar k}}\sin(\beta -2x_1). \eqno(9.6)
$$
Equation (9.3) may be then re-written as
\[
|T(k)|^2\tau_\Trm^\varphi (x_1,x_2;k)+  |R(k)|^2\tau_\Rrm^\varphi
(x_1,x_2;k) =
\]
$$
= \tau^\Drm(x_1,x_2;k) + \Delta\tau^\Drm(x<x_1;k) = \tau^\Drm(x_1,x_2;k) -
{\frac{mR}{\hbar k^2}}\sin(\beta -2kx_1)\,.\eqno(9.7)
$$
To support such a reasoning, those two authors stressed the fact that the
aforesaid self-interference term is completely independent of $T(k)$ and
of $\alpha (k)$, just because there is {\em not} interference for $x>d$.
Furthermore, they treated two particular cases as examples.

The first one regards an infinitely thick barrier (a step).
Being the barrier infinitely thick, there will be no transmitted particles
and all the particles will be reflected: $R=1$. It is therefore easy to
prove that
\[
\Delta \tau _{\Rrm}^{\varphi}={\frac{2}{\kappa v}}={\frac{2m}{\hbar k\kappa
}}\,,
\]
\[
\tau ^{\Drm}={\frac{E}{V_{0}}}\Delta \tau _{\Rrm}^{\varphi}\,,
\]
\[
\Delta \tau ^{\Drm}={\frac{E-V_{0}}{V_{0}}}\Delta \tau _{\Rrm}^{\varphi}\,.
\]
In this case no contrast arises between the extrapolated phase
time, which increases with $k^{-1}$ when $k\to 0$, and the dwell
time that, on the contrary, goes to 0 with $E/v_{\grm}\sim k$. In fact, if $
\tau ^{\Drm}$ is the time spent inside the barrier and $\Delta \tau ^{\Drm}$
the delay (or the advance) due to the self-interference, the latter term will be
the larger one. This because, the more the incident energy decreases, the less
the particle penetrates inside the barrier.

The second example concerns, instead, a Dirac $\delta$ barrier: a case that is
one of the first to be treated and solved in the literature about the
subject. Of course, in suc a case the dwell time has to be zero; nevertheless
we get:$^{\#9}$
\footnotetext{$^{\#9}$In the present case, $d$ has the role of a ``parameter''
only, since the Dirac $\delta $ is got as the limit of narrower and
narrower, and at the same time higher and higher, barriers (while the
area $V_{0}d$ is kept constant).}
\[
\Delta \tau _{\Rrm}^{\varphi}=\Delta \tau _{\Trm}^{\varphi}=T(k){\frac{
V_{0}d}{mv^{3}}}
\]
with:
\[
T(k)={\frac{1}{{1+\displaystyle{\left({\frac{V_{0}d}{\hbar k}}\right)^{2}}}}} \ .
\]
This means that in this case the tunnelling time can originate only from the
self-interference term.

\

\section{Generalization of the dwell time} 

As already said, not all the authors agree about the importance 
attributed till now to the dwell time, also [but not only] because of
eq.(1.2). In fact such
a relation, claimed to be a consequence of the superposition principle and of
the ``complementarity'' of transmission and reflection, would imply that
$$
\int_{{\rm Barrier}}|\psi(x,k)|^{2}dx = j\left(|T(K)|^{2}\tau_{\Trm} +
|R(k)|^{2}\tau_{\Rrm}\right)\,,
\eqno(10.1)
$$
which does a priori require at least that $\tau _{\Trm}=\tau _{\Rrm}$,
independently of the form of the potential barrier.

Besides that, eq.(1.1) had been obtained by B\"{u}ttiker[21] in 1983 following
a method which raised some critical comments[5,30,31]: In fact, even if
expressed in terms of $\psi(x,t)$, such a definition (1.1) does not appear to
fully account for the time evolution of the wavepacket. Moreover, except
for relation (1.2), the mentioned definition does not
suggest any method for distinguishing among the times corresponding to the
various processes entering the play. For this reason, trying to discriminate
among these times, Olkhovsky and Recami were initially induced to propose
--at a preliminary level-- the following
definitions for the transmission and reflection times:
$$
\overline{\tau_{\Trm}}=\overline{t(x_\frm)}_{\Trm}^{\IIIrm}-\overline{
t(x_\irm)}_{\inrm} =
{\frac{{\int_{-\infty}^{\infty} \drm t \, t\, J_{\Trm}^{\IIIrm}(x_\frm,t)}}{{
\int_{-\infty}^{\infty} \drm t \, J_{\Trm}^{\IIIrm}(x_\frm,t)}}} - {\frac{
\int_{-\infty}^{\infty} \drm t\, t\, J_{\inrm}(x_\irm,t)}{
\int_{-\infty}^{\infty} \drm t\, J_{\inrm}(x_\irm,t)}} =
$$
$$
= {\frac{\int_0^\infty \drm E \, v|g(E)T|^2 \tau_{\Trm}^{Ph} (x_{\irm},x_{\frm
};E)}{ \int_{0}^{\infty} \drm E\, v|g(E)T|^2}}
= (x_{\frm} -x_{\irm})\lan v^{-1}\ran_\Trm+\lan\delta\tau_\Trm\ran_\Trm\,,\eqno(10.2a)
$$
and
\[
\overline{\tau_\Rrm}=\overline{t(x_\frm)}_\Rrm^\Irm-\overline{t(x_\irm)}_{
\inrm} =
{\frac{{\int_{-\infty}^{\infty} \drm t \, t\, J_\Rrm^\Irm(x_\frm,t)}}{{
\int_{-\infty}^{\infty} \drm t \, J_\Rrm^\Irm(x_\frm,t)}}} - {\frac{
\int_{-\infty}^{\infty} \drm t\, t\, J_{\inrm}(x_\irm,t)}{
\int_{-\infty}^{\infty} \drm t\, J_{\inrm}(x_\irm,t)}} =
\]
\[
= {\frac{\int_0^\infty \drm E \, v|g(E)R|^2 \tau_{\Rrm}^{Ph} (x_{\irm},x_{\frm
};E)}{ \int_{0}^{\infty} \drm E\, v|g(E)R|^2}} =
(x_{\frm} -x_{\irm})\lan v^{-1}\ran_\Rrm+\lan\delta\tau_\Rrm\ran_\Rrm\,.\eqno(10.2b)
$$
In fact, since $J(x,t)\drm t$ represents the probability density for a particle
to pass through the point $x$ during the time interval $(t,t+\drm t)$, in order
to determine the average time at which a wavepacket $\Psi (x,t)$ reaches
the point $x$ we have to perform a weighted average over the variable $t$
by means of
$$
w(x,t) ={\frac{J(x,t)}{\int_{-\infty}^\infty J(x,t) \drm t}}. \eqno(10.3)
$$
Soon after, however, the same authors[5] noticed definitions (10.2) to
hold only
when the incident and transmitted wavepackets are totally separated
both in space and in time. Indeed, when $x_\irm$ and $x_\frm$ are not far
enough from the barrier walls, it is possible to have interference effects
between the incident and the reflected part. Moreover, {\em the sign} of the
current density $J(x,t)$ can change during the time evolution
of the packet (for instance when the peak of the incident wave reaches the
front-edge of the barrier). As a consequence, the integral \ $\int_{-\infty}^{\infty}\drm t\,t
J(x,t)$, \ which represents the algebraic sum of positive and negative
quantities (fluxes), as well as the probability densities $w(x,t)$, may be no longer
positive-definite quantities: And each probability density would be
endowed with a physical meaning only during the time intervals in which
the relevant current does not invert its direction. Therefore, it appears
necessary to break the mentioned integral into several integrals, each of them
taken over a time interval during which the sign of $J(x,t)$ is only positive
or only negative. In such a way we'll obtain probability densities everywhere
positive-definite:
\[
w_+(x,t)={\frac{J_{+}(x,t) \drm t}{\int_{-\infty}^{\infty} \drm t J_{+}(x,t)}
} \,, \qquad\quad w_-(x,t)={\frac{J_{-}(x,t) \drm t}{\int_{-\infty}^{\infty}
\drm t J_{-}(x,t)}} \ ,
\]
where $J_{+}$ and $J_{-}$ represent the positive and negative values of
$J(x,t)$, respectively. Taking such considerations into account, Olkhovsky
and Recami[5] were led to propose as average transmission and reflection times
the new expressions
$$
\overline{\tau_\Trm}=\overline{t(x_\frm)}_+ -\overline{t(x_\irm)}_+ = {\frac{
\int_{-\infty}^\infty \drm t \, t J_+(x_\frm,t)}{ \int_{-\infty}^\infty \drm
t\, J_+(x_\frm,t)}} - {\frac{\int_{-\infty}^\infty \drm t\, t J_+(x_\irm,t)}{
\int_{-\infty}^\infty \drm t\, J_{+}(x_\irm,t)}} \eqno(10.4a)
$$
and
$$
\overline{\tau_\Rrm}=\overline{t(x_\irm)}_- - \overline{t(x_\irm)}_+ = {
\frac{\int_{-\infty}^\infty \drm t\, t J_{-}(x_\irm,t)}{ \int_{-\infty}^
\infty \drm t\, J_{-}(x_\irm,t)}} - {\frac{\int_{-\infty}^\infty \drm t\, t
J_{+}(x_\irm,t)}{ \int_{-\infty}^\infty \drm t\, J_{+}(x_\irm,t)}}\,.
\eqno(10.4b)
$$
Before going on, it is important to stress that, starting only from the
continuity equation
\[
{\frac{\partial\rho (x,t)}{\partial t}}+\, {\frac{\partial J(x,t)}{\partial x
}}\, =0
\]
and from the standard quantum-mechanical probabilistic interpretation of
$\rho(x,t)$, one can easily {\em prove} that the above quantities
$w_{\pm}(x,t)$ correspond just to
the probability that our particle (moving forwards or coming backwards,
respectively) is located at point $x$ during the time interval
$(t,t+ \drm t)$. Actually, during each time interval in which it is
either $J=J_+$ or $J=J_-$, we can apply the continuity equation [which
is always valid] to $J_{\pm}$:
$$
{\frac{\partial\rho_{{}_>}(x,t)}{\partial t}}=  -{\frac{\partial J_+(x,t)}{
\partial x}} \eqno(10.5a)
$$
$$
{\frac{\partial\rho_{{}_<}(x,t)}{\partial t}}=  -{\frac{\partial J_-(x,t)}{
\partial x}} \ ,\eqno(10.5b)
$$
obtaining in such a way the two quantities $\partial\rho_{{}_>}(x,t)/\partial t$
and $\partial\rho_{{}_<}(x,t)/\partial t$. \ On integrating with respect
to time over the interval $(-\infty, t)$, we can then define:
$$
\rho_{{}_>}(x,t)=-\int_{-\infty}^t{\frac{\partial J_+(x,t)}{\partial x}}
\drm t^\prime\,,\eqno(10.6a)
$$
$$
\rho_{{}_<}(x,t)=-\int_{-\infty}^t{\frac{\partial J_-(x,t)}{\partial x}}
\drm t^\prime\,.\eqno(10.6b)
$$
Let us also impose the constraints \ $\rho_{{}_>}(x,-\infty )=0$ and 
$\rho_{{}_<}(x,-\infty )=0$: Let us suppose, in other words, that
initially the particle (or wavepacket) is infinitely far from $x$. By
integrating now with respect to $x$
we obtain two more quantities which we'll call $N_{{}_>}(x,\infty
;t)$, \ $N_{{}_<}(-\infty ,x;t)$ and for which one has
$$
N_{{}_>}(x,\infty ;t)=\int_x^\infty \rho_{{}_>}(x^\prime ,t)dx^\prime =
\int_{-\infty}^t J_+(x,t^\prime ) \drm t^\prime >0,\eqno(10.7a)
$$
$$
N_{{}_<}(-\infty ,x;t)= \int_{-\infty}^x \rho_{{}_<}(x^\prime ,t)
\drm x^\prime = - \,
\int_{-\infty}^t J_-(x,t^\prime ) \drm t^\prime >0\,.\eqno(10.7b)
$$
The last two expressions will give us the probability ---as a function of the
current densities $J_{\pm} (x,t)$--- that our particle, moving forward or
backwards, be located at time $t$ on the right or on the left of $x$,
respectively. Let us notice that the constraints $\rho_{{}_>}(x,-\infty)=0$, \
$\rho_{{}_<}(x,-\infty)=0$, which we imposed before integrating, are
equivalent now to \ \mbox{$J_{\pm}(-\infty,t)=0$.}
Finally, on differentiating again eq.(10.6) (now with respect to time), one
gets
$$
J_+(x,t)= {\frac{\partial}{\partial t}} N_{{}_>}(x,\infty ;t) > 0 \, , \eqno
(10.8a)
$$
$$
J_-(x,t)= {\frac{\partial}{\partial t}} N_{{}_<}(-\infty ,x;t) > 0 \, , \eqno
(10.8b)
$$
and then
$$
w_+(x,t) ={\frac{\displaystyle{\frac{\partial}{\partial t}}N_{{}_>}(x,\infty
;t)}{N_{{}_>}(x,-\infty ,\infty)}}\,,\eqno(10.9a)
$$
$$
w_-(x,t) ={\frac{\displaystyle{\frac{\partial}{\partial t}}N_{{}_<}(-\infty
,x;t)}{N_{{}_<}(-\infty ,x,\infty)}}\,.\eqno(10.9b)
$$
Such relations are sufficient fo justifying the quantum-mechanical probabilistic
interpretation of $w_+(x,t)$ and $w_{-}(x,t)$.

At this point we can define the {\em average value} of the time at which
the particle is at the point $x$, while moving in the positive or negative
direction along the chosen axis:$^{\#10}$
$$
\overline{t_{+}(x)}\ =\ {\frac{\int_{-\infty}^{\infty}t\,J_{+}(x,t)\drm t}{
\int_{-\infty}^{\infty}J_{+}(x,t)\,\drm t}}\,,\eqno(10.10a)
$$
$$
\overline{t_{-}(x)}\ =\ {\frac{\int_{-\infty}^{\infty}t\,J_{-}(x,t)\drm t}{
\int_{-\infty}^{\infty}J_{-}(x,t)\,\drm t}}\,.\eqno(10.10b)
$$
\footnotetext{$^{\#10}$We choose to call $x$ the propagation axis. However,
when discussing approaches different from ours, we maintain the notations adopted
by the original author (especially when the text is accompanied by a
figure taken from, or prompted by, the original paper).}
We are endowed now with all the means needed to define even the {\em variances}
of the distributions related to the above-mentioned times:
$$
\sigma ^{2}(t_{+}(x))={\frac{\int_{-\infty}^{\infty}t^{2}J_{+}(x,t)\drm t}{
\int_{-\infty}^{\infty}J_{+}(x,t)\drm t}}-\overline{(t_{+}(x))^{2}}\,,\eqno
(10.11a)
$$
$$
\sigma ^{2}(t_{-}(x))={\frac{\int_{-\infty}^{\infty}t^{2}J_{-}(x,t)\drm t}{
\int_{-\infty}^{\infty}J_{-}(x,t)\drm t}}-\overline{(t_{-}(x))^{2}}\,.\eqno
(10.11b)
$$
We then succeeded in constructing a formalism which allows us obtaining
both the average values and the variances (and other possible higher-order
moments) for the ``time distributions'' of all
the possible processes relevant to one-dimensional tunnelling. The same
definitions, anyway, can be extended to any other collision processes, even
different from tunnelling and in the presence of any kind of potentials.
As we have already seen, for the tunnelling and reflection times one has
$$
\overline{\tau _{\Trm}}(x_{\irm},x_{\frm})=\overline{t(x_{\frm})}_{+}-
\overline{t(x_{\irm})}_{+}=\eqno(10.4a)
$$
with $-\infty<x_{\irm}<0$ and $d<x_{\frm}<\infty$ and, according to eq.(10.11),
$$
\sigma ^{2}(\tau _{\Trm}(x_{\irm},x_{\frm}))=\sigma ^{2}(t_{+}(x_{\frm
}))+\sigma ^{2}(t_{+}(x_{\irm})).\eqno(10.12)
$$
Moreover, in the case $x_{\irm}=0$, $x_{\frm}=d$, we may write
$$
\cases{\overline{\tau_{{\rm Tun}}}(0,d)=\overline{t(d)}_+ -\overline{t(0)}_+\cr
       \sigma^2(\tau_\Trm(0,d))=\sigma^2(t_+(d)) +\sigma^2(t_+(0)).        \cr}
\eqno(10.13)
$$
Taking as an example $x_{\irm}=0$ and $0<x_{\frm}<d$, we can obtain
the penetration times inside the barrier as
$$
\overline{\tau _{{\rm Pen}}}(0,x_{\frm})=\overline{t(x_{\frm})}_{+}-
\overline{t(0)}_{+}\eqno(10.14)
$$
or, analogously, for $0<x<d$, the times
$$
\overline{\tau _{{\rm Ret}}}(x,x)=\overline{t(x)}_{-}-\overline{t(x)}_{+}
\eqno(10.15)
$$
while, for $-\infty <x_{\irm}<d$, it will be
$$
\overline{\tau _{\Rrm}}(X_{\irm},x_{\irm})=\overline{t(x_{\irm})}_{+}-
\overline{t(X_{\irm})}_{+}.\eqno(10.16)
$$
At last, let us re-examine, on the basis of the definitions reported above,
the previous definitions of phase time and dwell time. As far as the former
is concerned, it appears once again its merely asymptotic character. Indeed,
being the phase time deduced within to an explicitly stationary context,
on the basis of our previous results it can get a physical meaning only when
$x_{\irm}\to\infty $; that is, when $J_{+}(x,t)$ is the current
density of the initial packet in the absence of {\em any} interference
(between the transmitted and the reflected part) due to reflected
waves. Similarly, the dwell time,
represented by the equivalent expression[32,33]
\[
\overline{\tau}^{{\rm D}}(x_{\irm},x_{\frm})=\left[ \int_{-\infty}^{\infty
}t\;J(x_{\frm},t)\;\drm t-\int_{-\infty}^{\infty}t\;J(x_{\irm},t)\;\drm t\;
\right] \;\left[ \int_{-\infty}^{\infty}J_{\inrm}(x_{\irm},t)\;\drm t\,
\right] ^{-1} \ ,
\]
with $-\infty <x_{\irm}<0$, and $x_{\frm}>d$, is not, in general, physically
meaningful: In fact, the weight in the time averages is positive-definite,
and normalized to 1, only in the rare cases in which $x_{\irm}\to-\infty $
and $J_{\inrm}=J_{\IIIrm}$ \ (i.e., when the barrier is ``transparent").

\

\section{Penetration and return times: numerical results} 

The penetration and return times will be extensively analyzed elsewhere[8,31];
here they will be studied only briefly.

Equations (10.3) and (10.12-15) do not allow an easy analytical evaluation of
expressions for the tunnelling times (namely for penetration, return and
reflection), not even in the rather simple case of a square barrier. Let us
then present the results of numerical calculations for the average duration
of several penetration and return processes {\em for gaussian packets inside a
square barrier}, performed by Olkhovsky et al. in the second one of refs.[5].
Such calculations confirmed the existence of the Hartman effect, and seem
to be in agreement (due to theoretical connection between tunnelling and
evanescent-wave propagation) with the experimental data of Cologne, Berkeley,
Florence, Vienna, Orsay, Rennes, etc.

Let us remember that, following the previous notations, it holds
\[
\Psi _{\inrm}(x,t)=\int_{0}^{\infty}Cf(k-\overline{k})\ \exp [ikx-iEt/\hbar
]\;\drm k \ ,
\]
where
\[
f(k-\bar{k}) = \exp\left[{-{\frac{(k-\bar{k})^{2}}{2(\Delta k)^{2}}}}\right] \ ,
\]
$E=\hbar ^{2}k^{2}/2m$, \ $C$ is a normalization constant, and $m$ is, in this
case, the electron mass. The penetration lengths will be expressed in
angstroms, and the penetration times in seconds.

In Fig.6a the plots are shown of $\overline{\tau_{{\rm Pen}}}(0,x)$, with
$0<x<d$, corresponding to $d = 5 \;${\AA}, \ for \ $\Delta k =
0.02${\AA}$^{-1}$ and $ 0.01 \; {\rm {\AA}}^{-1}$. \ Notice right now that the
penetration time $\overline{\tau_{{\rm Pen}}}(0,x)$ does always show a clear
tendence to saturation. In Fig.6b we depict, instead, the plot corresponding to $d =
10\ ${\AA} and $\Delta k = 0.01 \; {\rm {\AA}}^{-1}$. It is interesting to
observe that, for constant $\Delta k$, the values of the {\em total}
penetration time $\overline{\tau_{{\rm Pen}}} \equiv
\overline{\tau_{{\rm Pen}}}(0,d)$
remain practically unchanged, when going on from $d = 5${\AA} to $d = 10 \;${\AA}:
a result which brings more evidence, once again, in support of the so-called Hartman
effect. Analogous results have been obtained even for $d > 10 \; {\rm {\AA}}$,
by varying the parameter $\Delta k$ between $0.005$ and $0.050 \;${\AA}$^{-1}$, and the
energy $\overline E$ in the range $1$ to $10\;$eV.

\

\noindent In Figs.\,7, 8 and 9 the behaviour is shown of the average
durations of
the penetration and return processes, as a function of the penetration length
(with $x_{\irm}=0$, and $0\leq \,x_{\frm} \equiv x \leq d$), for barriers
with height $V_{0}=10\;$eV, and width $d=5\;${\AA} or in some cases
$10\;${\AA}. \ In particular:\hfill\break
--- In Fig.7, the plots are presented of $\overline{\tau
_{{\rm Pen}}}(0,x)$, corresponding to different values of the average kinetic
energy: $\overline{E} = 2.5,\ 5$ and $7.5\;$eV with $\Delta k=0.02{\rm {\AA
}}$ (lines 1, 2 and 3); and $\overline{E}=5\;$eV with $\Delta k=0.04{\rm {\AA
}}^{-1}$ (line 4). In all the four cases it is $d=5\;{\rm {\AA}}$.\hfill\break
--- In Fig.8 we show the plots of $ \overline{\tau _{{\rm
Pen}}}(0,x)$ corresponding to: $d=5\;{\rm {\AA}}$, with $\Delta k=0.024$ and
$0.04\;{\rm {\AA}}^{-1}$ (lines 1 and 2); and to $d=10\;{\rm {\AA
}}$, with $\Delta k=0.02$ and $0.04\;{\rm {\AA}}^{-1}$ (lines 3 and 4). The
average kinetic energy $\overline{E}$ is $5\;$eV, i.e., half the
barrier energy $V_{0}=10\;$eV.\hfill \break
--- Finally, in Fig.9 we present some plots of $\overline{\tau}_{{\rm
Ret}}(x,x)$. The lines 1, 2 and 3 correspond to: 
$\overline{E}= 2.5,\ 5$ and $7.5\;$eV, respectively, with $\Delta k=
0.02\;{\rm {\AA}}^{-1}$ and $d=5\;{\rm {\AA}}$. The lines 4, 5 and 6 correspond,
instead, to: $\overline{E}= 2.5,\ 5$ and $7.5\;$eV,
with $\Delta k=0.04\;{\rm {\AA}}^{-1}$ and $d=5\;{\rm {\AA}}$; while the lines 7 and 8
correspond to \ \ $\Delta k=0.02$
and $0.04\;{\rm {\AA}}^{-1}$, respectively, 
with $\overline{E}=5$ eV, and, this time, $d=10 {\rm {\AA}}$.

\

\noindent With regard to the employed numerical methods, those authors
remarked the integration in d$t$ to have been performed using the time
interval $[-10^{-13},\ +10^{-13}]\;$s, symmetric with respect to $t=0$:
An interval three orders of magnitude larger than the temporal
width of the wavepacket,
which, in its turn, is of the order of $1/(\bar{v}\,{\Delta k})=(
{\Delta k}\,\sqrt{2
\overline{E}/m} )^{-1} \sim 10^{-16}\;$s.  This is
equivalent to consider the evolution of the wavepacket along 
the time interval
[$-\infty ,\;\;+\infty $], without assigning to it any finite
starting time $t$; in agreement with the relations
$J_{\pm}(-\infty ,t)=0$ and, equivalently, with $\rho _{{}_{<}}(x,-\infty )=0$.
Let us moreover remark that the considered
packet has been constructed in order that its centroid arrives at $x_{0}$ at
time $t=0$.

\

\noindent From the above Figures 7)--9) it can be inferred that:

1) the average duration of a tunnelling process $\overline{\tau_{{\rm Tun}}}
(0,d)$ does not depend on the width $d$ of the barrier (Hartman effect);

2) the penetration time increases quickly only at the
beginning of the barrier, namely, in the barrier region close to $x= 0$;

3) $\overline{\tau_{{\rm Pen}}}(0,x)$ tends to a saturation value in the
final part of the barrier, that is, for $x\to d$.

\

\noindent With regard to the effects mentioned at points 1)-3) above, they
could be caused, according to the same authors, by interference {\em inside
the barrier} between the forward propagating [sometimes called, loosely
speaking, "incoming" or "entering" or penetrating]
and the backwards propagating ["returning" or reflected] waves, whose
superposition produces the $J_{+}$ and $J_{-}$ fluxes. See
also the pictures (in particular Fig.3, p.351) in the second one of
refs.[5]. \

Eventually, in connection with the plots of
$\overline{\tau _{{\rm Ret}}}(x,x)$ as a function of $x$, shown in
Fig.9, we notice that:

4) the average duration of the reflection, $\overline{\tau_{\Rrm}}(0,0) \equiv
\overline{\tau_{{\rm Ret}}}(0,0)$, {\em does not depend} on the barrier
width $d$;

5) between $0$ and $\sim 0.6\,d$ the value of
$\overline{\tau_{{\rm Ret}}}(0,x)$ is approximately constant;

6) the value of $\overline{\tau_{{\rm Ret}}}(0,x)$ increases with $x$ only
in the region close to $x=d$ (even if, it was stressed by the mentioned authors,
the calculations in such a region are not very accurate, because
quantity $\int_{-\infty}^{\infty} J_{-}(x,t) \drm t$ takes on very small
values in correspondence to it).

Notice that point 4) ---a result expected, like point 1), for quasi-monochromatic
particles also on the basis of Dumont and Marchioro's paper[34]--- agrees with the
data obtained by Steinberg et al.[35] for arbitrary wavepackets. As before,
also points 5) and 6) can be due to interference phenomena inside the
barrier. Actually, if the returning wavepacket is almost totally damped
down near $x=d$ by the penetrating wavepacket, only a negligible part of its
back-tail (composed by the slower components) will survive.
When $x$ decreases ($x\to 0$), the non-damped part of the returning
packet is larger (including the faster components), so
that the difference $\overline{\tau _{{\rm Ret}}}(0,x)-\overline{\tau _{{\rm
Pen}}}(0,x)$ remains approximately constant. Moreover, the interference
between the incident and the reflected waves in the region $x\leq 0$
causes $\overline{t_{-}(0)}$ to be larger than
$\overline{t_{+}(0)}$: 
This could explain, going back to our terminology, the larger values of
$\overline{\tau_{\Rrm}}(0,0)$ with respect to
$\overline{\tau_{{\rm Tun}}}(0,d)$. \ Let us recall, and stress, that the
mentioned interference between reflected and incoming waves does cause an
{\em acceleration} of the wavepacket peak, so that in general its speed
becomes Superluminal even before (near) the barrier.

\

\section{On the general validity of the Hartman effect} 

Let us recall that we named Hartman Effect (HE) the independence of the mean
tunnelling time from the barrier width, so that for large (opaque) barriers the
effective tunnelling--velocity can become arbitrarily large. \ Such effect has
been analyzed in the first part of Sect.\,3.
Now we shall briefly discuss the validity of the HE for all the other theoretical
expressions proposed for the mean tunnelling times. \ Let us first consider
the {\em mean dwell time}, the {\em mean Larmor time},     
and {\em the real part} of the complex tunnelling time obtained by averaging
over the Feynman paths.   
All of them, in the case of quasi-monochromatic particles and opaque
rectangular barriers, become equal to $\hbar k/(\kappa V_{0})$: \ And one
immediately verifies$^{\#11}$ that also for all such mean tunnelling times
\footnotetext{$^{\#11}$See refs.[47] below, besides refs.[31].}
there is {\em no} dependence on the barrier width, and
consequently the HE is valid. \ The validity of the HE for the mean tunnelling
time has been proved in 1992, as we know, in the nonrelativistic approach
by Olkhovsky and Recami, an approach developed in refs.[5,30,31] and
moreover confirmed by the numerical simulations performed in the same
set of papers (for various cases of gaussian wavepackets).

By contrast, the ``second Larmor time'' $\tau_{z\Trm}^\Lrm$, as well as
the B\"{u}ttiker-Landauer time $\tau_\Trm^{\BLrm}$  
and the imaginary part $\Imrm \ \tau_\Trm^\Omega$  
of the complex tunnelling time obtained within the Feynman approach, which too are equal to
$\tau_{z\Trm}^\Lrm$, yield the result \ $md/(\hbar\kappa)$ in the opaque
rectangular-barrier limit[30,31]: That is, they all are {\em proportional}
to the barrier width $d$, so that the HE is {\em not} valid for them! \
However, it has been shown in refs.[5,8] that$^{\#12}$ such last three times 
\footnotetext{$^{\#12}$See also refs.[47], below.}
{\em are not mean times}, but rather {\em standard deviations} (or ``mean
square fluctuations") of the tunnelling-time distributions,
In conclusion, the latter three times are not connected with the peak
(or group) velocity of the tunnelling particles, but with the {\em spread} of
the tunnelling velocity distributions.

All the results have been obtained for transparent media (without absorption
or dissipation). As it was theoretically demonstrated in ref.[36] within
nonrelativistic quantum mechanics, the HE vanishes for barriers with high
enough absorption. This was confirmed experimentally for electromagnetic
(microwave) tunnelling in ref.[37].

Let only add a comment. From some recent papers[38], it seems that the
integral penetration time, needed to cross a portion of a barrier, in the
case of a very long barrier starts to increase again ---after the plateau
corresponding to infinite speed--- proportionally to the distance. This is
due to the contribution of the above-barrier frequencies
contained in the considered wavepackets, which become more and more
important as the tunnelling components are progressively damped down. \ In
this paper, however, we refer to the behaviour of the {\em tunnelling} (or,
in the classical case, of the evanescent) waves only.

Actual deviations from the ``Hartman effect"-behaviour, however, exist.
They are very interesting: but will be considered in ref.[8], and elsewhere;
here, we shall just mention some of them in our Sect.14.

\section{Optical equivalence of tunnelling} 

As already mentioned, several experimental evidences of the Hartman effect
have been obtained during the last ten years, or so, in a series of measurements made in
Cologne,[39]
Berkeley,[40] Florence,[41], Vienna[42], Rennes, Orsay, etc. However, those
measurements regarded the transmission times of electromagnetic waves (including
optival photons) 
by examining the propagation of evanescent modes inside a ``classical
barrier", like a segment of waveguide for ``below cut-off"
frequencies, or ``frustrated refraction'' regions, respectively.
Vrious experiments employing quantum particles, such as, e.g., electrons, have
been also proposed: But such measurements seem to be still difficult to
perform, particularly because of the very small times involved in the
tunnelling processes. For a Josephson junction, for instance, such times
result to be of about 10 fs, and maybe of the order of 1 fs in other
solid-state devices. For optical systems, instead, such times are already of
the order of some ps, for frequencies in the visible range, and reach the ns
in some of the experiments made in Florence and Cologne with microwaves.

Even without reviewing here such celebrated[43] experiments, and the
results by them obtained, let us however recall --in connection with
them-- the equivalence between the {\bf classical} transmission of evanescent
electromagnetic modes and the {\em quantum} tunnelling of
particles. As an example of systems with classical barriers, we slall fix
our attention on waveguides. \ Let us consider a particle with mass $m$ and
kinetic
energy $E=\hbar ^{2}k^{2}/2m$. For (one-dimensional) quantum propagation
in the presence of a uniform potential $V_{0}$, the {\em Schr\"{o}dinger
equation} for such a
particle writes:
$$
{\frac{\partial ^{2}\psi}{\partial x}}+{\frac{2m}{\hbar ^{2}}}(E-V_{0})\psi
=0.\eqno(12.1)
$$
If we define
$$
\kappa ^{2}\equiv {\frac{2m}{\hbar ^{2}}}(E-V_{0}),\eqno(12.2)
$$
Equation (12.1) results to be formally identical to the {\em Helmholtz equation} for
any (electric or magnetic) component of an e.m. field propagating through a
dispersive medium:
$$
{\frac{\partial^{2}\psi}{\partial x}}+\kappa^{2}\psi = 0\,.\eqno(12.3) \ ,
$$
with
\[
\kappa ={\frac{2\pi}{\lambda_{m}}}={\frac{2\pi}{\lambda}}n,
\]
$\lambda _{m}\;$ being the wave-length inside the medium, \ $\lambda \;$ the
wave-length in vacuum, \ and $n$ the refraction index of the medium in which
the field propagates. The comparison between the two equations suggests the
obvious correspondence:
\[
\sqrt{{\frac{2m}{\hbar ^{2}}}(E-V_{0})}\to{\frac{2\pi}{\lambda}}n.
\]
For a ``square" waveguide, with dimensions $a{\times}b$ ($a<b$), and with
perfectly conducting walls, we know that:\
$$
\kappa ={\frac{2\pi}{\lambda}}\sqrt{1-\left( {\frac{\lambda}{2b}}\right)
^{2}}={\frac{2\pi}{\lambda}}\sqrt{1-\left( {\frac{\lambda}{\lambda _{c}}}%
\right) ^{2}}\,,\eqno(12.4)
$$
where $\lambda _{c}=2b$ is the ``cut-off" wave-length above which the
square-root term becomes negative and, as a consequence, $\kappa $ turns out
to be imaginary. Since $\lambda =c/\nu =2\pi c/\omega $, one has
$$
\kappa =\sqrt{{\frac{\omega ^{2}}{c^{2}}}-{\frac{\pi ^{2}}{b^{2}}}}={\frac{%
\omega}{c}}\sqrt{1-\left( {\frac{\omega _{c}}{\omega}}\right) ^{2}}\,,
\eqno(12.5)
$$
with $\omega _{c}=\pi c/b$ ($\nu _{c}=c/2b$), that yields, in its turn, the
cut-off frequency below which $\kappa $ becomes {\em imaginary} [as it is known
to happen for a tunnelling particle too].

Notice that the dispersion relation for a square waveguide is surprisingly
similar[44] (as noticed also in the ``Feynman Lectures"[7]) to the one of a
relativistic particle, once we make the substitution
\[
{\frac{\pi}{b}}={\frac{\omega_c}{c}}= {\frac{mc}{\hbar}}.
\]
In fact, on multiplying eq.(12.5) by $\hbar$, we obtain
\[
\hbar\kappa =\sqrt{{\frac{(\hbar\omega)^2}{c^2}} -{\frac{(2\hbar\pi )^2}{b^2}%
}}\to p^2=\sqrt{{\frac{E^2}{c^2}}-mc^2} \ .
\]
Let us emphasize that a comparison is being made
between the Helmholtz equation, which is {\em relativistic} and {\em %
classical} [it is obtained from Maxwell equations), and the
Schr\"{o}dinger's one, which is a {\em non-relativistic} and {\em quantum}
equation. \ Let us remark that, on the basis of the
Maxwell equations only, computer simulations have been carried on[45],
which numerically verified the Superluminality of the evanescent waves along
``under-sized'' (sub-critical) waveguides; further computers
simulations have been performed more extensively[46], with analogous
results.

Let us go back to our eqs.(12.1) and
(12.3). By differentiating eqs.(12.2) and (12.5), we obtain
\[
v_{{\rm group}}^{{\rm particle}}={\frac{\drm\omega}{\drm\kappa}}={\frac{%
\hbar}{m}}\kappa \ ; \qquad \quad v_{{\rm group}}^{{\rm e.m.wave}}={\frac{\drm%
\omega}{\drm\kappa}}={\frac{c^{2}}{\omega}}\kappa .
\]
Through the replacement
$$
{\frac{\hbar}{m}}\;\to\;{\frac{c^{2}}{\omega}}={\frac{c}{2\pi \nu
}} \ , \eqno(12.6)
$$
one can observe a mathematical identity between the electromagnetic results,
for transmission of microwaves
through a waveguide, and the quantum-theoretical predictions for 
one-dimensional motion of a particle in the presence of a uniform potential.$^{\#13}$
\footnotetext{$^{\#13}$Notice that eq.(12.6) is equivalent to the relation $%
\hbar \omega =mc^{2}$.} Actually, in both cases the solutions of the two
equations are given by linear combinations of the wave-functions
\[
\psi (x,t)=e^{{\pm}i\kappa x}e^{i\omega t} \ .
\]
In particular, as already mentioned, when the particle energy results
{\em smaller} than $V_{0}$ or when, in
electromagnetic case, the angular frequency $\omega $ becomes {\em smaller}
than $\omega _{c}$, quantity $\kappa $ becomes imaginary and the wavepacket
exponentially decrease as ${\rm e}^{-|\kappa |x}$ (tunnelling, or evanescent
wave propagation, respectively).
Notice that the field (or, analogously, the particle wave-function) penetrate
in a similar way along an undersized segment, i.e., the forbidden region, of
the waveguide (or inside the quantum barrier, for particles), up to a
distance of the order of $|\kappa |^{-1}$.
Obviously, despite of the formal analogies, there are physical differences
between the tunnelling of electrons and the propagation of microwaves below
cut-off in
waveguides. In fact, as already said, the Helmholtz wave-equation and
the Schr\"{o}dinger particle-equation are mathematicaaly the same equation;
but, whilst in the first case what propagates is a field component
(it is the field itself), in the second case, instead, it is the particle
wave-function. Nevertheless, since in both cases we have to do with the
time-evolution of {\em wavepackets}, nothing prevents us from interpreting
the results of experiments with electromagnetic waves (microwaves, e.g.)
as ``physical simulations'' of electron tunnelling.  The
results of such ``simulations''[39-43] reproduce quite well also the
quantum-theoretical predictions, so that the equivalence between the
two cases is verified.

More subtle is the circumstance that in the {\em time-dependent} case the
Schroedinger equation (in the presence of barrier) and the Helmholtz equation
(for electromagnetic waves in a waveguide) are no more mathematically
equivalent, since the time-derivative is of the first order in the former
case and of the second order in the latter case. Anyway, it can be
seen[47,31,8]
that those equations still admit analogous classes of solutions, different
only for their ``spreading'' properties.

Notice explicitely that all the same can be repeated, of course, for
other classical barriers, as any band-gap filters (cf., e.g.,
refs.[35,40-42]).

\

We cannot skip mentioning, at last, the surprising results of one-dimensional
non-resonant tunnelling through {\em two} successive opaque potential
barriers (see the last four refs.[31]), separated
by an intermediate free region $\Rcal$, by analyzing the relevant solutions
to the Schroedinger equation. In Olkhovsky, Recami and Salesi[31], it has
been found that the total traversal time
does {\em not} depend not only on the opaque barrier widths (``Hartman
effects"), but also on the width $R$ of region $\Rcal$: so that the effective
velocity in
the region $\Rcal$, between the two barriers, can be regarded as infinite. \
This agrees with the results known from the corresponding metallic waveguide
experiments[39], which simulated the mentioned
tunnelling experiment just because of the known mathematical identity between Schroedinger and
Helmholtz equation.  It is worth mentioning that the above prediction of
quantum mechanics have been theoretically
confirmed and generalized (and explained in terms of ``superoscillations")
by Aharonov et al.[31]: Indeed, the claim of those authors is that, according
to QM, a wavepacket can travel, in zero time and negligible distortion, a distance
arbitrarily larger than the width of the wavepacket. From the experimental
point of view, Olkhovsky, Recami and Salesi's[31] prediction has been re-verified by
Longhi et al.[31] on using as (classical) barriers two gratings in an
optical fiber.  We shall spend a few more words about these question in the
Appendix.

\section{A brief mention of some experimental results} 

As we have seen, the problem stated at the beginning ---how much time does a
particle spend to cross a potential barrier--- even though
substantially solved, still remains debated. On the one hand, it actually
exists experimental evidence supporting, e.g., the simple definition of phase
time. On the other hand, this very definition, and the same experimental
results, force us to accept, in certain cases, the arising of group-velocities
larger than the speed of light in vacuum (Hartman effect, as it was called
by Olkhovsky and Recami). It
must be noticed however that the appearance of Superluminal, or even
of {\em negative} group-velocities in classical optics is {\em not} a new
phenomenon: It was studied and experimentally observed in
works[48,49] that only recently received the attention they deserved. As a
matter of fact, in a {\em dispersive linear non-absorbing} medium
\[
v_\grm=\displaystyle{\frac{\drm\omega}{\drm k}}= \displaystyle{\frac{c}{
n(\omega )+ \omega n^\prime (\omega)}}\ ,
\]
and in regions with strong {\em anomalous dispersion} (near a resonance), the
group-velocity can exceed $c$ or even, as already said, become negative.
Various authors have proven or proposed that in those cases the
group-velocity may lose its
ordinary physical meaning, so that no signal (no information) can indeed
be transmitted by the medium with velocity
larger than $c$ [we are referring ourselves to the mathematical formulations,
and interpretations, in Chap.7 of
Jackson's {\em Classical Electrodynamics}, or in Chap.3 of Sommerfeld's book].
In those cases, moreover, the phenomenon appears to be 
easily explained in terms of the reshaping produced by the
attenuation of the less energetic, and slower, components.
The outgoing signal, therefore, may seem to have travelled at velocities
larger than the velocity at which its more energetic components did
actually travel.

In the tunnelling case, however, we have seen in Sect.3 that (both for particles and for
e.m. waves) we can always be able to avoid the effects due to possible reshaping,
or to the transmission of initially faster particles. This is experimentally
supported also from the fact that, as it has been seen, e.g., in the
experiments by Enders and Nimtz, the width of the ``signals" remains
unaffected (something relevant when thinkink of a transmission by Morse's
alphabet), even if their amplitude decreases. \
One has to notice that the barrier-crossing of particles is a
statistical process, in the sense that one cannot know a priori which
particle will pass through the barrier.
This is true, of course; but the
weight of such a consideration becomes lower when it grows the number of the 
particles at our disposal for attempting a ``signal" transmission: To remain
within the Morse alphabet example, one can send out dots and dashes by
emitting pulses of, say, one thousand and ten thousand particles each,
respectively; the dot and dashes will then be recognized also after
the tunnelling. This becomes even more meaningful when one approaches the
classical limit. \  The claim that Superluminal tunnelling 
cannot be used to transmit any information is in need, therefore, for
further discussion; more details can be found in refs.[50,51].

In tunnelling, moreover, the appearance of Superluminal group-velocities
takes place in general when the probability associated with such an
event is rather low.  This circumstance is at the basis of the
the interpretation put forward by Steinberg et
al.[10]. According to them, in the Superluminal tunnelling processes the
corresponding crossing times should be regarded as {\em weak values},
coming from {\em weak measurements}. The concept of weak measure had been
introduced in 1988 by Aharonov et al.[52] starting from the 
``classical" measurement theory by von Neumann[53].  According to the authors
of ref.[10], when we make a measurement on a subset associated with
low probability (as, e.g., the one composed by the transmitted particles),
and this subset belongs to a set on which it has been
made a weak measurement (that is, a measurement with a large uncertainty,
that leaves quasi-undisturbed the whole set), it is possible to obtain
as the result of the measurement on the subset a value completely different
from all the eigenvalues accessible to the system. Such a value would not
be however a value really assumed by the system, corresponding to a
wave-function that is not an eigenstate.

According to other authors, on the contrary, the Superluminal
velocities associated with tunnelling would be actually real; while what
is to be re-interpreted is the causality principle.[54-56,50] \ To this purpose,
let us remember that Special Relativity can be extended ---without dropping the
ordinary Postulates--- so as to include Superluminal motions. Such an
``extended relativity''[55], in other words, can incorporate tachyons
without destroying Einstein's relativity, but only extending it to the
new velocity realm. In particular, it is possible to solve the causal
problems and the so-called causal paradoxes.[56,50] \ It is also worthwhile
recalling that extended relativity itself predicts, 
on the basis of simple classico-geometrical considerations,
the transition for any Superluminal object (wave pulse, or particle)
from positive to negative group-velocities (whenever the object ``overcomes"
the infinite speed): see refs.[55,57]. The fact that to negative speeds
there correspond negative crossing times[57], becomes interesting in the
light of many subsequent, more or less recent, experimental results[58].

We shall come back to such problems in the Appendix, where we present a
bird's-eye view of all the sectors of physics in which Superluminal motions
seem to appear (cf. also ref.[59]).

\

\

{\bf ACKNOWLEDGEMENTS}

The Authors are grateful to F.Bassani, A.Paoletti, R.A.Ricci and to C.Vasini
for stimulating discussions and their kind interest.  Thanks are also due
for scientific collaboration to A.\,Agresti, V.\,Abate, M.\,Baldo,
J.D.\,Bekenstein, N.\,Ben-Amots, A.\,Bertin, G.\,Bonera, R.\,Bonifacio,
L.\,Bosi, M.\,Brambilla, G.\,Brown, D.\,Campbell, G.\,Cavalleri, R.\,Chiao,
C.\,Cocca, C.\,Conti, C.A.Dartora, G.\,Degli Antoni,
S.\,Esposito, J.R.\,Fanchi, F.\,Fontana, R.\,Garavaglia, A.\,Gigli Berzolari,
L.\,Horwitz, H.E.\,Hern\'{a}ndez, L.C.\,Kretly, G.\,Kurizki, J.\,Jakiel,
J.\,-y.Lu, G.D.\,Maccarrone, A.\,van der Merwe, D.\,Mugnai, G.\,Nimtz,
K.Z.\,N\'obrega, V.\,Petrillo, M.\,Pernici, A.\,Ranfagni, F.\,Raciti, B.Reznik,
A.\,Shaarawi, P.\,Saari, D.\,Stauffer, A.M.\,Steinberg, M.T.\,Vasconselos,
M.\,Villa, A.\,Vitale, A.K.\,Zaichenko and M.\,Zamboni-Rached.

\newpage

{\bf Captions of the Figures (of the Text)}

\

Fig.1 -- One-dimensional tunnelling of a particle through a rectangular
potential barrier of height $V_0$ and width $d$. (The propagation axis
is considered to be the $x$-axis).

\

Fig.2 -- Time evolution of a minimum-uncertainty wavepacket (at $t=0$)
incident on an infinitely high rectangular
barrier which extends on the right of $z=0$. The initial width and centroid
of the packet are 3${\rm {\AA}}$ and at $-20\;{\rm {\AA}}$, respectively. The
average energy of the incident packet is 4 eV. (In this figure,
the propagation axis has been called $z$).

\

Fig.3 -- Transmission coefficients $T(k) \equiv A(k)$ calculated
for different values of the thickness $d$ of a barrier consisting in a square
potential with height $V_0=\hbar^2\varepsilon^2/2m$. In this figure it
is moreover presented the plot of $f(k-k_0)=e^{-(k-k_0)^2/2(\Delta k)^2}$, with $
k_0=0.7\varepsilon$ and $\Delta k=0.1k_0$.

\

Fig.4 -- Plots of $T(k)f(k-k_0)$, for below-barrier energies,
as a function of the wave-number $k$, for different
values of $k_0$ and different values of the barrier thickness $d \equiv a$ \
(Fig.4a: $k_0=0.3\;\varepsilon$; \ Fig.4b: $k_0=0.5\;\varepsilon$; \ $Fig.4c:
k_0=0.7\;\varepsilon$; \ Fig.4d: $k_0=0.9\;\varepsilon$), with \
$\Delta k=0.1 \,k_0$; \ and, from up to down: \ $a=1/\varepsilon$, $a=3/\varepsilon$, $
a=6/\varepsilon$, $a=10/\varepsilon$, $a=15/\varepsilon$, $a=20/\varepsilon$,
$a=25/\varepsilon$. \ Let us recall (see the text) that $\varepsilon \equiv
2mV_{0}/\hbar $. \ Notice that the line corresponding to the last
value does not appear in Fig.4a, because of the strong
attenuation of the peak.

\

Fig.5 -- The quantum clock of Baz' and Rybachenko: a particle entering
the barrier starts to suffer a Larmor precession in the presence of a magnetic field
confined inside the barrier. The spin of a tunnelling particle is turned
parallell to the direction of the field. \ Notice that, following the
original paper, in this figure the propagation axis has been called $y$.

\

Figs.6 -- Fig.\,a): Behaviour of the average penetration time,
$\overline{\tau_{{\rm Pen}}}(0,x)$ (expressed in
seconds), as a function of the penetration length $x_{ \frm} = x$ (expressed
in angstrom) for a square barrier with width $d = 5 \; {\rm {\AA}}$, with
$\Delta k = 0.02 \; {\rm {\AA}}^{-1}$ (broken line) \ and \ $\Delta k = 0.01
\; {\rm {\AA}}^{-1}$ (continuous line). \ It is worth noticing that
$\overline{\tau_{{\rm Pen}}}(0,x)$
increases quickly for the first initial angstroms ($\sim 2.5$ {\AA}), and
afterwards approaches a ``saturation'' value. This supports the
existence of the Hartman effect. \ Fig.\,b): As in 6a), with $\Delta k = 0.01 \;
{\rm {\AA}}^{-1}$, but with a barrier endowed with a double thickness, $d = 10
\; {\rm {\AA}}$ . \ Notice that the numerical values of the {\em total}
tunnelling time $\overline{\tau_{{\rm Tun}}}(0,d)$ remain practically
unchanged when we go on from \ $d = 5
\;${\AA} \ to \ $d = 10 \;${\AA}, \ as a further evidence of the appearance
of the Hartman effect.

\

Fig.7 -- Behaviour of $\overline{\tau_{{\rm Pen}}}(0,x)$ (expressed in seconds) as
a function of $x$ (in angstroms), relative to tunnelling through a
barrier of thickness $d = 5 \;${\AA} and for several values of ${\overline E}
$ and of $\Delta k$: \ \ line 1: \ $\Delta k = 0.02 {\rm {\AA}}^{-1}$ and $
\overline E = 2.5 \; {\rm eV}$; \ \ line 2: \ $\Delta k = 0.02 {\rm {\AA}}
^{-1}$ and $\overline E = 5.0 \; {\rm eV}$; \ \ line 3: \ $\Delta k = 0.02
{\rm {\AA}}^{-1}$ and $\overline E = 7.5 \; {\rm eV}$; \ \ line 4: \ $\Delta
k = 0.04 {\rm {\AA}}^{-1}$ and $\overline E = 5.0 \; {\rm eV}$.

\

Fig.8 -- Behaviour of $\overline{\tau_{{\rm Pen}}}(0,x)$ (in seconds) as a
function of $x$ (in angstroms) for $\overline E = 5 \; {\rm eV}$, and
for different values of $d$ and $\Delta k$: \ \ line 1: \ $d = 5 \; {\rm {\AA
}}$, $\Delta k = 0.02 {\rm {\AA}}^{-1}$; \ \ line 2: \ $d = 5 \; {\rm {\AA}}$
, $\Delta k = 0.04 {\rm {\AA}}^{-1}$; \ \ line 3: \ $d =10 \; {\rm {\AA}}$, $
\Delta k = 0.02 {\rm {\AA}}^{-1}$; \ \ line 4: \ $d =10 \; {\rm {\AA}}$, $
\Delta k = 0.04 {\rm {\AA}}^{-1}$.

\

Fig.9 -- Behaviour of $\overline{\tau_{{\rm Ret}}}(x,x)$ (in seconds) as a
function of $x$ (in angstroms) for different values of $d$, $\overline E$
and $\Delta k$; \ namely:
line 1: \ $d = 5 \; {\rm \AA}$, \ $\overline E = 2.5 \; {\rm eV}$ and
$\Delta k = 0.02 {\rm \AA}^{-1}$; \ \
line 2: \ $d = 5 \; {\rm \AA}$, \ $\overline E = 5.0 \; {\rm eV}$ and
$\Delta k = 0.02 {\rm \AA}^{-1}$; \ \
line 3: \ $d = 5 \; {\rm \AA}$, \ $\overline E = 7.5 \; {\rm eV}$ and
$\Delta k = 0.02 {\rm \AA}^{-1}$; \ \
line 4: \ $d = 5 \; {\rm \AA}$, \ $\overline E = 2.5 \; {\rm eV}$ and
$\Delta k = 0.04 {\rm \AA}^{-1}$; \ \
line 5: \ $d = 5 \; {\rm \AA}$, \ $\overline E = 5.0 \; {\rm eV}$ and
$\Delta k = 0.04 {\rm \AA}^{-1}$; \ \
line 6: \ $d = 5 \; {\rm \AA}$, \ $\overline E = 7.5 \; {\rm eV}$ and
$\Delta k = 0.04 {\rm \AA}^{-1}$; \ \
line 7: \ $d = 10 \; {\rm \AA}$, \ $\overline E = 5.0 \; {\rm eV}$ and
$\Delta k = 0.02 {\rm \AA}^{-1}$; \ \
line 8: \ $d = 10 \; {\rm \AA}$, \ $\overline E = 5.0 \; {\rm eV}$ and
$\Delta k = 0.04 {\rm \AA}^{-1}$.

\newpage

\centerline{\bf REFERENCES (of the Text)}

\

\noi [1] E.U. Condon: Rev. Mod. Phys. {\bf 3}, 43 (1931).

\noi [2] L.A. MacColl: Phys. Rev. {\bf 40}, 621 (1932).

\noi [3] V.S. Olkhovsky, E. Recami \& A.I. Gerasimchuk: Nuovo Cimento {\bf A22},
263 (1974); \ E. Recami: ``A time operator and the time--energy uncertainty
relation", in {\em The Uncertainty Principle and Foundations of Quantum
Mechanics}, ed. by W.C.Price \& S.S.Chissick (J.Wiley; London, 1977), p.21;
E. Recami: ``An operator for Time", in {\em Proceedings of the Karpacz
Winter School (Recent Developments in Relativistic QFT and its Application,
vol. 2)}, ed. by W.Karwowski (Wroclaw Univ. Press; Wroclaw, 1976), p.251,
and refs. therein; \ V.S. Olkhovsky: Sov. J. Part. Nucl. {\bf 15}, 130 (1984);
 \ Nukleonika {\bf 35}, 99 (1990); \ ``The study of nuclear reactions
by their temporal properties", in {\em Nuclear Reaction Mechanisms}, ed. by
D.Seeliger \& H.Kalka (World Scientific; Singapore, 1991), p.15.

\noi [4] S. Collins, D. Lowe \& J.R. Barker: J. Phys. {\bf C20}, 6233 (1989);
\ E.H. Hauge \& J. A. St\o vneng: Rev. Mod. Phys. {\bf 61}, 917 (1989); \ A.P.
Jauho: ``Tunnelling times in heterosctructures: A critical review", in {\em
Hot Carriers in Semiconductor Nanostructures: Physics and Applications}
(A.T.T. Company; 1992) pp.121-150.

\noi [5] V.S. Olkhovsky \& E. Recami: Phys. Reports {\bf 214}, 339 (1992); \
V.S. Olkhovsky, R. Recami, F. Raciti \& A.K. Zaichenko: J. de Physique-I {\bf
5}, 1351 (1995). \ See also Ref.[8] below.

\noi [6] R. Landauer \& Th. Martin: Rev. Mod. Phys. {\bf 66}, 217 (1994).

\noi [7] R.P. Feynman, R.B. Leighton \& M. Sands: {\em The Feynman lectures on
Physics}, (Addison-Wesley; 1977), vol.2, pp.24-27.

\noi [8] V.S.Olkhovsky, E.Recami \& J.Jakiel: ``Unified time analysis of
photon and nonrelativistic particle tunnelling" [Lanl Archives e-print
\# quant-ph/0102007], to appear in Physics Reports.

\noi [9] T.E. Hartman: J. Appl. Phys. {\bf 33}, 3427 (1962); \
J.R. Fletcher: J. Phys. {\bf C18}, 155 (1985). \ See also A.
Tagliacozzo: Nuovo Cim. {\bf D10}, 363 (1988); A. Tagliacozzo \& E. Tosatti:
Phys. Scripta {\bf 38}, 301 (1988).

\noi [10] See for instance A.M. Steinberg: J. Physique {\bf I4}, 1813
(1994), and refs. therein; \ Phys. Rev. {\bf A52}, 32 (1995). \ Cf. also
K. Hauss \& P. Busch: Phys. Lett. {\bf A185}, 9 (1994).

\noi [11] E.H. Hauge, J.P. Falck \& T.A. Fjeldly: Phys. Rev. {\bf B36}, 4203
(1987).

\noi [12] C.R. Leavens \& G.C. Aers: Phys. Rev. {\bf B89}, 1202 (1989).

\noi [13] Th. Martin \& R. Landauer: Phys. Rev. {\bf A47}, 2023 (1993).

\noi [14] M. B\"{u}ttiker \& R. Landauer: Phys. Rev. Lett. {\bf 49}, 1739
(1982).

\noi [15] M. B\"{u}ttiker \& R. Landauer: Phys. Scr. {\bf 32}, 429 (1985)

\noi [16] M. B\"{u}ttiker \& R. Landauer: I.B.M. J. Res. Dev. {\bf 30}, 451
(1986).

\noi [17] A.J. Baz': Sov. J. Nucl. Phys. {\bf 4}, 182 (1967).

\noi [18] A.J. Baz': Sov. J. Nucl. Phys. {\bf 5}, 161 (1967).

\noi [19] V.E. Rybachenko: Sov. J. Nucl. Phys. {\bf 5}, 635 (1967).

\noi [20] J.J. Sakurai: {\em Meccanica Quantistica Moderna}, (Zanichelli,
1990) pp.75-77.

\noi [21] M. B\"{u}ttiker: Phys. Rev. {\bf B27}, 6178 (1983).

\noi [22] J.P. Falck \& E.H. Hauge: Phys. Rev. {\bf B38}, 3287 (1988).

\noi [23] M.J. Hagmann: Solid State Commun. {\bf 82}, 867 (1992).

\noi [24] D. Sokolovski \& L.M. Baskin: Phys. Rev. {\bf A36}, 4604 (1987).

\noi [25] P. H\"{a}nggi: In {\em Lectures on Path Integration} (World
Scientific; London, 1991), pp.352.

\noi [26] P. Sokolovski \& J.N.L. Connor: Phys. Rev. {\bf A47}, 4667 (1993). \
See also H.A. Fertig: Phys. Rev. Lett. {\bf 65}, 2321 (1990); Phys. Rev. {\bf
B47}, 1346 (1993).

\noi [27] C.R. Leavens \& G.C. Aers: In {\em Scanning Tunnelling Microscopy --
III}, ed. by R. Wiesedanger \& H.J. G\"{u}ntherodt (Springer; New York, 1993),
p.105.

\noi [28] E. Madelung: Z. Phys. {\bf 40}, 332 (1926); \ G. Salesi: Mod. Phys.
Lett. {\bf A11} (1996) 1815; \ G. Salesi \& E. Recami: Found. Phys. Lett.
{\bf 10} (1997) 533; \ E. Recami \& G. Salesi: Phys. Rev. {\bf A57} (1998)
98; \ Found. Phys. {\bf 28} (1998) 763-776; \ K. Imafuku, I. Ohba \& Y.
Yamanaka: Phys. Rev. {\bf A56}, 1142 (1997); \ M. Abolhasani \& M. Golshani:
Phys. Rev. {\bf A62} (2000), issue June 14; \ I.L. Egusquiza \& J.G. Muga:
Phys. Rev. {\bf A61} (1999), issue December 9; \ J. Le\`{o}n, J. Julve, P.
Pitanga \& F.J. de Urr\`{i}es: e-print quant-ph/9903060; \ J. Le\`{o}n: J.
Phys. {\bf A40}, 4791 (1997).

\noi [29] F.T. Smith: Phys. Rev. {\bf 118}, 349 (1960).

\noi [30] V.S. Olkhovsky, E. Recami \& A.K. Zaichenko: Solid State Commun.
{\bf 89}, 31 (1994).

\noi [31] A.P.L.Barbero, H.E.Hern\'andez F., \& E.Recami: Phys. Rev.
{\bf E62}, 8628 (2000) and refs. therein; \ V.S.Olkhovsky, E.Recami \&
J.Jakiel: ref.[8]; \ J.
Jakiel, V.S. Olkhovsky \& E. Recami: Physics Letters {\bf A248} 156, (1998);
\ V.S. Olkhovsky, R. Recami, F. Raciti \& A.K. Zaichenko: ref.[5]. \ See
also V.S. Olkhovsky, R. Recami \& G. Salesi: Europhy. Lett. {\bf 57},
879 (2002); \ Y. Aharonov, N. Erez \& B. Reznik: Phys. Rev.{\bf A65},
no.052124 (2002); \ S. Longhi, P. Laporta, M. Belmonte \& E. Recami:``Measurement
of Superluminal optical tunneling times in double-barrier photonic bandgaps",
Phys. Rev.{\bf E65}, no.046610 (2002); \ and \ S. Esposito: Phys. Rev.
{\bf E67}, no.016609 (2003).

\noi [32] W. Jaworski \& D.M. Wordlawd: Phys. Rev. {\bf A37}, 2834 (1998).

\noi [33] C.R. Leavens: Solid State Com. {\bf 85}, 115 (1993).

\noi [34] R.S. Dumont \& T.L. Marchioro: Phys. Rev. {\bf A47}, 85 (1993).

\noi [35] A.M. Steinberg, P.G. Kwiat \& R.Y. Chiao: Phys. Rev. Lett. {\bf 71},
708 (1993).

\noi [36] F. Raciti \& G. Salesi: J. Phys. (France) {\bf I4}, 1783 (1994).
\ See, however, also S. Baskoutas, A. Jannussis \& R. Mignani: J. Phys. {\bf
A27}, 2189 (1994).

\noi [37] G.Nimtz, H.Spieker \& M.Brodowsky: J. Phys. (France) {\bf I4}, 1379
(1994).

\noi [38] M.Abolhasani \& M.Golshani: ref.[28]; \ J.Le\`on: (private
commun.); \ V.Petrillo \& L.Refaldi: ``A time asymptotic expression for the
wave-function emerging from a quanto-mechanical barrier", sub. for pub.; \
L.Refaldi: M.Sc. Thesis (V.Petrillo, R.Bonifacio, E.Recami supervisors), Phys.
Dept., Milan University, 2000.

\noi [39] A. Enders \& G. Nimtz: J. de Physique {\bf I2}, (1992) 1693; {\bf
3}, 1089 (1993);  Phys. Rev. {\bf B47}, 9605 (1993);  Phys. Rev. {\bf E48} 632
(1993); \ G. Nimtz, A. Enders \& H. Spieker: J. de Physique {\bf I4}, 1
(1994); W. Heitmann \& G. Nimtz: Phys. Lett. {\bf A196}, 154 (1994); \ G.
Nimtz, A. Enders \& H. Spieker: in {\em Wave and Particle in Light and Matter}
(Proceedings of the Trani Workshop, Italy, Sept.1992), ed. by A. van der Merwe
\& A.Garuccio (Plenum; New York, in press); \ H. Aichmann \& G. Nimtz,
``Tunnelling of a FM-Signal: Mozart 40," submitted for pub.; \ G. Nimtz \& W.
Heitmann: Prog. Quant. Electr. {\bf 21}, 81 (1977).

\noi [40] A.M. Steinberg, P.G. Kwiat \& R.Y. Chiao: ref.[33]; \ R.Y.Chiao,
P.G. Kwiat \& A.M. Steinberger: Scientific American {\bf 269} (1993), issue
no.2, p.38; \ A.M. Steinberg \& R.Y. Chiao: Phys. Rev. {\bf A51}, 3525 (1995).
 \ Cf. also P.G. Kwiat, A.M. Steinberg, R.Y. Chiao, P.H. Eberhard \&
M.D. Petroff: Phys. Rev. {\bf A48}, R867 (1993); \ E.L. Bolda, R.Y. Chiao \&
J.C. Garrison: Phys. Rev. {\bf A48}, 3890 (1993); A.M. Steinberg: Phys. Rev.
Lett. {\bf 74}, 2405 (1995); \ R.Y. Chiao \& A.M. Steinberg: ``Tunnelling
times and Superluminality", in {\em Progress in Optics}, ed. by E.Wolf, vol.37
(1997).

\noi [41] A. Ranfagni, P. Fabeni, G.P. Pazzi \& D. Mugnai: Phys. Rev. {\bf E48},
1453 (1993).

\noi [42] Ch. Spielmann, R. Szipocs, A. Stingl \& F. Krausz: Phys. Rev. Lett.
{\bf 73}, 2308 (1994).

\noi [43] Scientific American: an article in the Aug. 1993 issue; Nature:
comment ``Light faster than light?" by R.Landauer, Oct. 21, 1993; New
Scientist: editorial ``Faster than Einstein" at p.3, plus an article by
J\.Brown at p.26, April 1995.

\noi [44] Th. Martin \& R. Landauer: Phys Rev. {\bf 45A}, 2611 (1992); \
R.Y. Chiao, P.G. Kwiat \& A.M. Steinberg: Physica {\bf B175}, 257 (1991); \ A.
Ranfagni, D. Mugnai, P. Fabeni \& G.P. Pazzi: Appl. Phys. Lett. {\bf 58}, 774
(1991); and refs. therein. \ See also A.M. Steinberg: {\bf Phys. Rev.}
{\bf A52}, 32 (1995).

\noi [45] H.M. Brodowsky, W. Heitmann \& G. Nimtz: Phys. Lett. {\bf A222},
125 (1996).

\noi [46] A.P.L. Barbero, H. Hern\'{a}ndez F., \& E. Recami: ref.[31].

\noi [47] V.S.Olkhovsky, E.Recami \& J.Jakiel: ref.[8]; \ J. Jakiel,
V.S. Olkhovsky \& E. Recami: ref.[31]; \ V.S. Olkhovsky \& A. Agresti: in
{\em tunnelling and its Applications} (World Sci.; Singapore, 1997), pp.327-355.

\noi [48] C.G.B. Garret \& D.E. McCumber: Phys. Rev. {\bf A}, 305 (1970).

\noi [49] S. Chu \& S. Wong: Phys. Rev. Lett. {\bf 49}, 1293 (1982). \ See
also B.Segard \& B.Macke: Phys. Lett. {\bf A109}, 213-216 (1985).

\noi [50] E.Recami, F.Fontana \& R.Garavaglia: Int. J. Mod. Phys. {\bf A15},
2793 (2000), where also a proper definition of group-velocity for
evanescent waves is given. \ See also D. Mugnai, A. Ranfagni,
R. Ruggeri, A. Agresti \& E. Recami: Phys. Lett. {\bf A209}, 227 (1995). 

\noi [51] Cf. R.Y.Chiao and A.M.Steinberg: in {\em Progress in Optics},
ed.by E.Wolf (Elsevier Science; Amsterdam, 1997), vol.37, pp.346-405; \
R.W.Ziolkowski: Phys. Rev. E63 (2001) 046604; \
A.M.Shaarawi and I.M.Besieris: J. Phys. A:Math.Gen. 33 (2000) 7227-7254;
7255-7263; \ P.W.Milonni: J. Phys. B35 (2002) R31-R56; \ as well as
G.Nimtz and W.Heitman: Progr. Quant. Electr. 21 (1997)
81-108; \ G.Nimtz and A.Haibel: Ann. Phys. (Leipzig) 11 (2002) 163-171; \
G.Nimtz: IEEE J. Select. Top. Quantum Electron. 9(1) (2993) 163-171.

\noi [52] Y. Aharonov, D.Z. Albert \& L. Vaidman; Phys. Rev. Lett. {\bf 60},
1351 (1988); \ Y. Aharonov \& L. Vaidman: Phys. Rev. {\bf A41}, 11 (1990).

\noi [53] J. von Neumann: {\em Mathematical Foundations of Quantum Mechanics}
(Princeton Univ. Press; Princeton, 1983).

\noi [54] I.M. Duck, P.M. Stevenson \& E.C.G. Sudarshan: Phys. Rev. {\bf D40},
40 (1989).

\noi [55] See, e.g., E. Recami: ``Classical tachyons and possible
applications," Rivista Nuovo Cim. 9 (1986), issue no.6, pp.1-178, and refs.
therein; \ E. Recami \& W.A. Rodrigues Jr.: ``A model theory for tachyons
in two dimensions", in {\em Gravitational Radiation and Relativity,} ed. by
J. Weber \& T.M. Karade (World Scient.; Singapore, 1985), pp.151-203.

\noi [56] E. Recami: ``Tachyon kinematics and causality", Foundation of
Physics {\bf 17}, 239 (1987); \ ``The Tolman `Anti-telephone' paradox:
Its solution by tachyon mechanics," Lett. Nuovo Cim. {\bf 44}, 587 (1985).

\noi [57] V.S.Olkhovsky, E.Recami, F.Raciti \& A.K.Zaichenko: ref.[5],
p.1361. \ See also refs.[53,54], and p.2807 in E.Recami, F.Fontana \& 
R.Garavaglia: ref.[55].

\noi [58] See, e.g., S.Longhi, M.Marano, P.Laporta, M.Belmonte \& P.Crespi:
Phys. Rev. {\bf E65} (2002), no.045602(R); \
L.J.Wang, A.Kuzmich \& A.Dogariu: Nature {\bf 406}, 277 (2000); \
G.Nimtz: Europ. Phys. J.-B (to appear as a Rapid Note); \
M.W.Mitchell \& R.Y.Chiao: Phys. Lett. {\bf A230}, 133-138 (1997); \
B.Segard \& B.Macke: ref.[49]; \
S.Chu \& W.Wong: ref.[49].

\noi [59] E. Recami: Found. Phys. {\bf 31} (2001) 1119-1135.

\newpage

\centerline{\large{\bf APPENDIX}}

\

{\bf A1. - Introduction.}\hfill\break

\hspace*{5ex} The question of Super-luminal ($V^{2}>c^{2}$) objects or waves
has a long story, starting perhaps in 50 b.C. with Lucretius' {\em De Rerum
Natura} (cf., e.g., book 4, line 201). \
Still in pre-relativistic times, one meets various related works, from those by
J.J.Thomson to the papers by the great A.Sommerfeld. \ With Special
Relativity, however, since 1905 the conviction spread over that the speed $c$
of light in vacuum was the {\em upper} limit of any possible speed. For
instance, R.C.Tolman in 1917 believed to have shown by his ``paradox'' that
the existence of particles endowed with speeds larger than $c$ would have
allowed sending information into the past. Such a conviction blocked for
more than half a century ---apart from an isolated paper (1922) by the
Italian mathematician G.Somigliana--- any research about Superluminal
speeds. Our problem started to be tackled again essentially in the fifties
and sixties, in particular after the papers[A1] by E.C.George Sudarshan et
al., and later on[A2] by E.Recami, R.Mignani, et al. [who rendered the
expressions subluminal and Superluminal of popular use by their works at the
beginning of the Seventies], as well as by H.C.Corben and others (to confine
ourselves to the {\em theoretical} researches). \ The first experiments
looking for tachyons were performed by T.Alv\"{a}ger et al.
For references, one can check pages 162-178
in ref.[A1], where about 600 citations are listed; pages 285-290 in ref.[A3];
pages 592-597 of ref.[A4] or pages 295-298 of ref.[A5]; as well as the large
bibliographies by V.F.Perepelitsa[A6] and as the book in ref.[A7]. \ In
particular, for the causality problems one can see refs.[A1,A8] and references
therein, while for a model theory for tachyons in two dimensions one can be
addressed to refs.[A1,A9]. \ The first experiments looking for tachyons were
performed by T.Alv\"{a}ger et al.; some citations about the early
experimental quest for Superluminal objects may be found e.g. in
refs.[A1,A10].\\

\hspace*{5ex} In recent years, terms as ``tachyon'' and ``Superluminal''
fell unhappily into the (cunning, rather than crazy) hands of
pranotherapists and mere cheats, who started squeezing money out of
simple-minded people; for instance by selling plasters (!) that should cure
various illnesses by ``emitting tachyons''... \ We are dealing with them
here, however, since at least four different experimental sectors of physics
seem to indicate the actual existence of Superluminal motions,
thus confirming some long-standing theoretical predictions[A3]. \
In this rapid informative Appendix, after a sketchy theoretical
introduction, we are going to set forth a reasoned outline of the experimental
state-of-the-art: brief, but accompanied by a bibliography sufficient in some
cases to provide the interested readers with coherent, adequate information;
and without forgetting to call attention ---at least in the two sectors more
in fashion today--- to some other worthy experiments.

\

{\bf A2. Special and Extended Relativity}.\hfill\break

\hspace*{5 ex} Let us state first that special relativity (SR), abundantly
verified by experience, can be built on two simple, natural Postulates: \ 1)
that the laws (of electromagnetism and mechanics) be valid not only for a
particular observer, but for the whole class of the ``inertial" observers: \
2) that space and time are homogeneous and space is moreover isotropic. \
From these Postulates one can theoretically deduce that one, and only
one, {\em invariant} speed exists: and experience tells us such a speed to be that,
$c$, of light in vacuum; in fact, light possesses the peculiar feature of
presenting always the same speed in vacuum, even when it runs towards or away
from the observer. \ It is just that feature, of being invariant, that makes quite
exceptional the speed $c\/$: no bradyons, and no tachyons, can enjoy the same
property.

\hspace*{5ex} Another (known) consequence of our Postulates is that the
total energy of an ordinary particle increases when its speed $v$ increases,
tending to infinity when $v$ tends to $c$. Therefore, infinite forces would
be needed for a bradyon to reach the speed $c$. This fact generated the
popular opinion that speed $c$ can be neither achieved nor overcome. \
However, as speed-$c$ photons exist, which are born live and die always at
the speed of light (without any need for accelerating from rest to the light
speed), so particles can exist ---tachyons[A4]--- always endowed with speeds $%
V$ larger than $c$ (see Fig.A1). \ This circumstance has been picturesquely
illustrated by George Sudarshan (1972) with reference to an imaginary
demographer studying the population patterns of the Indian subcontinent: $<<$%
Suppose a demographer calmly asserts that there are no people North of the
Himalayas, since none could climb over the mountain ranges! That would be an
absurd conclusion. People of central Asia are born there and live there:
they did not have to be born in India and cross the mountain range. So with
faster-than-light particles$>>$. \ Let us add that, still starting from the
above two Postulates (besides a third one, even more obvious), the theory of
relativity can be generalized[A3,A4] in such a way to accommodate also
Superluminal objects; such an extension is largely due to
a series of works performed mainly in the Sixties--Seventies. \
Also within such an ``Extended Relativity''[A3] the speed $c$, besides being
invariant, is a limiting velocity: but every limiting value has two sides,
and one can a priori approach it both from the left and from the right.

\hspace*{5ex} Actually, the ordinary formulation of SR is too much
restricted. For instance, {\em even leaving tachyons aside}, it can be easily so
widened as to include {\em antimatter\/}[A5]. Then, one finds space-time to
be a priori populated by normal particles P (which travel forward in time
carrying positive energy), {\em and} by dual particles Q ``which travel
backwards in time carrying negative energy''. The latter shall appear to us
as {\em antiparticles}, i.e., as particles ---regularly travelling forward
in time with positive energy, but--- with all their ``additive'' charges
(e.g., the electric charge) reversed in sign: see Fig.A2. \ To clarify this
point, let us recall that we, macroscopic observers, have to move in time
along a single, well-defined direction, to such an extent that we
{\em cannot} even see a motion backwards in time; and every object like Q,
travelling backwards in time (with negative energy), will be
{\em necessarily} reinterpreted by us as an anti-object, with opposite charges but
travelling forward in time (with positive energy).[A3-A5]

\hspace*{5 ex} But let us forget about antimatter and go back to tachyons. A
strong objection against their existence is based on the opinion that by
tachyons it should be possible to send signals into the past, owing to the fact
that a tachyon T which ---say--- appears to a first observer $O$ as emitted
by A and absorbed by B, can appear to a second observer $O^{\prime}$ as a
tachyon T' which travels backwards in time with negative energy. However, by
applying (as it is obligatory to do) the ``reinterpretation rule" or
switching procedure seen above, T' will appear to the new observer 
$O^{\prime}$ just as an antitachyon ${\overline{{\rm T}}}$ emitted by B and
absorbed by A, and therefore travelling forward in time, even if in the
contrary {\em space} direction. In such a way, every travel towards the
past, and every negative energy, do disappear.

Starting from this observation, it is possible to solve[A5] the so-called
causal paradoxes associated with Superluminal motions: paradoxes which
result to be the more instructive and amusing, the more sophisticated they
are; \ but that cannot be re-examined here (some of them having been
proposed by R.C.Tolman, J.Bell, F.A.E.Pirani, J.D.Edmonds and others).[A6,A3]
\ Let us only mention here the following. \ The reinterpretation principle
---according to which, in simple words, signals are carried only by objects
which appear to be endowed with positive energy--- does eliminate any
information transfer backwards in time, but this has a price: That of
abandoning the ingrained conviction that the judgement about what is cause
and what is effect is independent of the observer. In fact, in the case
examined above, the first observer $O$ considers the event at A to be the
cause of the event at B. \ By contrast, the second observer $O^{\prime}$
will consider the event at B as causing the event at A. \ All the observers
will however see the cause to happen chronologically {\em before} its own
effect.

\hspace*{5ex} Taking new objects or entities into consideration always
forces us to a criticism of our prejudices. If we require the phenomena to
obey the {\em law} of (retarded) causality with respect to all the
observers, then we cannot demand also the {\em description} ``details" of
the phenomenon to be invariant too: Namely, we cannot demand in that case also the
invariance of the ``cause'' and ``effect'' labels.[A6,A2] \ To illustrate the
nature of our difficulties in accepting that e.g. the parts of cause and
effect depend on the observer, let us cite an analogous situation that does
not imply present-day prejudices: $<<$For ancient Egyptians, who knew only
the Nile and its tributaries, which all flow South to North, the meaning of
the word ``south'' coincided with the one of ``upstream'', and the meaning
of the word ``north'' coincided with the one of ``downstream''. When
Egyptians discovered the Euphrates, which unfortunately happens to flow
North to South, they passed through such a crisis that it is mentioned in
the stele of Tuthmosis I, which tells us about {\em that inverted water that
goes downstream (i.e. towards the North) in going upstream}$>>$ (Csonka,
1970).

\hspace*{5ex} The last century theoretical physics led us in a natural way
to suppose the existence of various types of objects: magnetic monopoles,
quarks, strings, tachyons, besides black-holes: and various sectors of
physics could not go on without them, even if the existence of none of them
is certain (also because attention has not yet been paid to some links
existing among them: e.g., a Superluminal electric charge is expected to
behave as a magnetic monopole; and a black-hole a priori can be the source
of tachyonic matter). \ According to Democritus of Abdera, everything that
was thinkable without meeting contradictions had to exist somewhere in the
unlimited universe. This point of view ---which was given by M.Gell-Mann the
name of ``totalitarian principle''--- was later on expressed (T.H.White) in
the humorous form ``Anything not forbidden is compulsory''. Applying it to
tachyons, Sudarshan was led to claim that $<<$if tachyons exist, they must to be
found; if they do not exist, we must be able to say clearly why...$>>$\newline

\

{\bf A3. The experimental state-of-the-art}.\hfill\break

\hspace*{5 ex} Extended Relativity can allow a better understanding of many
aspects also of {\em ordinary} relativistic physics, even if tachyons would
not exist in our cosmos as asymptotically free objects. \ As already said,
we are dealing with them ---however--- since their topic is presently
returning in fashion, especially because of the fact that at least four
different experimental sectors of physics seem to suggest the
possible existence of faster-than-light motions. \ We wish to put forth in
the following some information (mainly bibliographical) about the
experimental results obtained in each one of those different physics sectors.%
\newline

\

{\bf A)} \ {\bf Neutrinos} -- First: A long series of experiments, started
in 1971, seems to show that the square ${m_{0}}^{2}$ of the mass $m_{0}$ of
muonic neutrinos, and more recently of electronic neutrinos too, is
negative; which, if confirmed, would mean that (when using the na\"{\i}ve
language, commonly adopted) such neutrinos possess an ``imaginary mass'' and
are therefore tachyonic, or mainly tachyonic.[A7,A3] \ [In Extended
Relativity, the dispersion relation for a free tachyon becomes \ $E^{2}-{%
\mbox{\boldmath $p$}}^{2}=-m_{{\rm o}}^{2}$, and there is {\em no} need
therefore for imaginary masses.].

\

{\bf B)} \ {\bf Galactic Micro-quasars} -- Second: As to the {\em apparent}
Superluminal expansion observed in the core of quasars[A8] and, recently, in
the so-called {\em galactic microquasars\/}[A9], we shall not deal here with that
problem, too far from the other topics of this paper: without mentioning that
for those astronomical observations here exist orthodox interpretations,
based on ref.[A10], that are accepted by the astrophysicists' majority (even
if hampered by statistical considerations).
\ For a theoretical discussion --considering all the possible explanations,
the ``Superluminal" ones included--, see ref.[A11]. Here, let us mention only that
simple geometrical considerations in Minkowski space show that a {\em single}
Superluminal light source would look[A11,A3]: \ (i) initially, in the
``optical boom'' phase (analogous to the acoustic ``boom'' produced by a plane
travelling with constant supersonic speed), as an intense source which
suddenly appears, and later on becomes weaker; and that \ (ii) afterwards
seems to split into TWO objects receding
one from the other with speed $\;V>2c$.

\

{\bf C)} \ {\bf Evanescent waves and ``tunnelling photons''} -- Third:
Within quantum mechanics (and precisely in the {\em tunnelling} processes),
it had been shown that the tunnelling time ---firstly evaluated as a simple
``phase time'' and later on calculated through the analysis of the
wavepacket behaviour--- does not depend on the barrier width in the case of
opaque barriers (``Hartman effect'')[A12]: which implies Superluminal and
arbitrarily large (group) velocities $V$ inside long enough barriers: see
Figs.6 of the text. \ Experiments that may verify this prediction by, say, electrons are
difficult. Luckily enough, however, the Schroedinger equation in the
presence of a potential barrier is mathematically identical to the Helmholtz
equation for an electromagnetic wave propagating, e.g., down a metallic
waveguide along the $x$-axis: and a barrier height $U$ bigger than the
electron energy $E$ corresponds (for a given wave frequency) to a waveguide
transverse size lower than a cut-off value. A segment of undersized guide
does therefore behave as a barrier for the wave (photonic barrier)[A13]: So
that the wave assumes therein ---like an electron inside a quantum
barrier--- an imaginary momentum or wave-number and gets, as a consequence,
exponentially damped along $x$. In other words, it becomes an {\em evanescent%
} wave (going back to normal propagation, even if with reduced amplitude,
when the narrowing ends and the guide returns to its initial transverse
size). \ Thus, a tunnelling experiment can be simulated[A13] by having
recourse to evanescent waves (for which the concept of group velocity can be
properly extended[A14]). And the fact that evanescent waves travel with
Superluminal speeds has been actually {\em verified} in a series of famous
experiments (cf. Fig.A3).

\hspace*{5 ex} Namely, various experiments ---performed since 1992 onwards
by G.Nimtz at Cologne[A15], by R.Chiao's and A.Steinberg's group at
Berkeley[A16], by A.Ranfagni and colleagues at Florence[A17], and by others at
Vienna, Orsay, Rennes[A17]--- verified that ``tunnelling photons" travel with
Superluminal group velocities. Such experiments roused a great deal of
interest[A18], also within the non-specialized press, and were reported by
Scientific American, Nature, New Scientist, and even Newsweek, etc. \ Let us
add that also Extended Relativity had predicted[A19] evanescent waves to be
endowed with faster-than-$c$ speeds; the whole matter appears to be
therefore theoretically selfconsistent. \ The debate in the current
literature does not refer to the experimental results (which can be
correctly reproduced by numerical elaborations[A20,A21] based on Maxwell
equations only), but rather to the question whether they allow, or do not
allow, sending signals or information with Superluminal speed[A21,A14].

\hspace*{5 ex} Let emphasize that the {\em most interesting} experiment of
this series is the one with two ``barriers" (e.g., with two segments of
udersized waveguide separated by a piece of normal-sized waveguide: Fig.A4).
For suitable frequency bands ---i.e., for ``tunnelling" far from
resonances---, it was found that the total crossing time does not depend on
the length of the intermediate (normal) guide: namely, that the beam speed
along it is infinite[A22]. \ This agrees with what predicted by Quantum
Mechanics for the non-resonant tunnelling trough two successive opaque
barriers (the tunnelling phase time, which depends on the entering energy,
has been shown by us to be {\em independent} of the distance between the two
barriers[A23]).  Let us here repeat that the above prediction of
quantum mechanics, even if rather surprising, has been theoretically
confirmed and generalized by Aharonov et al.[A23]: Indeed, those
authors have found that, according to QM, a wavepacket can travel,
in zero time and negligible distortion, a distance
arbitrarily larger than the width of the wavepacket. From the experimental
point of view, our prediction[A23] has been re-verified by
Longhi et al.[A23] on using as (classical) barriers two gratings in an
optical fiber. Such important experiments could and should be repeated,
taking advantage also of the circumstance that quite interesting evanescence
regions can be easily constructed in the most varied manners, by
several ``photonic band-gap" filters (including photonic crystals). In
any case, both this ``extended Hartman effect", and the Hartman effect
itself, can result to be important for {\em applications} ---various
of which can be easily imagined---, even more than for theory. 

\hspace*{5ex} We cannot skip a further topic ---which, being delicate,
should not appear in a brief review like this one--- since one of the very last
experimental contribution to it (performed at Princeton in 2000 by J.Wang
et al.) raised a lot of interest. \
Even if in Extended Relativity all the ordinary causal paradoxes seem to
be solvable[A3,A6], nevertheless one has to bear in mind that (whenever it is
met an object, ${\cal O}$, travelling with Superluminal speed) one may have
to deal with {\em negative contributions} to the tunnelling times[A24]: and this
ought not to be regarded as unphysical. In fact, whenever an ``object''
(particle, electromagnetic pulse) ${\cal O}$ {\em overcomes} the
infinite speed[A3,A6] with respect to a certain observer, it will afterwards
appear to the same observer as the ``{\em anti}-object'' $\overline{{\cal O}}
$ travelling in the opposite {\em space} direction[A3,A6]. \ For instance,
when going on from the lab to a frame ${\cal F}$ moving in the {\em same}
direction as the particles or waves entering the barrier region, the object $%
{\cal O}$ penetrating through the final part of the barrier (with almost
infinite speed[A12,A21,A23], like in Figs.6 of the text) will appear in the frame ${\cal F}$
as an anti-object $\overline{{\cal O}}$ crossing that portion of the barrier
{\em in the opposite space--direction\/}[A3,A6]. In the new frame ${\cal F}$,
therefore, such anti-object $\overline{{\cal O}}$ would yield a {\em negative%
} contribution to the tunnelling {\em time\/}: which could even result, in total, to
be negative. \ For any clarification or explanation, see refs.[A18]. \ 
What we want to stress here is that the appearance of such negative times
is predicted by Relativity itself, on the basis of the ordinary
postulates[A3,A6,A24,A12,A21]. \ (In the case of a non-polarized beam,, the
anti-packet coincides with the initial wave packet; if a photon is however
endowed with helicity $\lambda =+1$, the anti-photon will bear the opposite
helicity $\lambda =-1$). \ From the theoretical point of view, besides
refs.[A24,A12,A21,A6,A3], see refs.[A25]. \ On the (quite interesting!)
experimental side, see papers [A26], the last one having already been
mentioned above.

\hspace*{5ex} Let us {\em add} here that, via quantum interference effects
in three-levels atomic systems, it is possible to obtain dielectrics with
refraction indices very rapidly varying as a function of frequency, with
almost complete absence of light absorption (i.e., with quantum induced
transparency) [A27]. \ The group velocity of a light pulse propagating in
such a medium can decrease to very low values, either positive or negatives,
with {\em no} pulse distortion. \ It is known that experiments were
performed both in atomic samples at room temperature, and in Bose-Einstein
condensates, which showed the possibility of reducing the speed of light to
few meters per second. \ Similar, but negative group velocities ---implying
a propagation with Superluminal speeds thousands of time higher than the
previously mentioned ones--- have been recently predicted, in the presence
of such an ``electromagnetically induced transparency'', for light moving in
a rubidium condensate[A28], while corresponding experiments are being
done, e.g., at the Florence European laboratory ``LENS''.

\hspace*{5 ex} Finally, let us emphasize that faster-than-$c$ propagation of
light pulses can be (and was, in same cases) observed also by taking
advantage of anomalous dispersion near an absorbing line, or nonlinear and
linear gain lines, or nondispersive dielectric media, or inverted two-level
media, as well as of some parametric processes in nonlinear optics (cf.
G.Kurizki et al.'s work).

\

{\bf D)} \ {\bf Superluminal Localized Solutions (SLS) to the wave
equations. The ``X-shaped waves"} -- The fourth sector (to leave aside the
others) is not less important. It returned in fashion when some groups of
scholars in engineering (for sociological reasons, most physicists
had abandoned the field) rediscovered by a series of works that any
wave equation ---to fix the ideas, let us think of the electromagnetic
case--- admit also solutions so much sub-luminal as Super-luminal (besides
the ordinary waves endowed with speed $c/n$). \ Let us recall that,
starting with the pioneering work by H.Bateman, it had slowly become known
that all wave equations (in a general sense: scalar,
electromagnetic, spinorial) admit wavelet-type solutions with
sub-luminal group velocities[A29]. \ Subsequently, also Superluminal
solutions started to be written down, in refs.[A30] and, independently, in
refs.[A31] (in one case just by the mere application of a Superluminal
Lorentz ``transformation"[A3,A32]).

\hspace*{5ex} A quite important feature of some new solutions of these is
that they propagate as localized, non-diffractive pulses: namely, according
to the Courant and Hilbert's[A29] terminology, as ``undistorted progressive
waves''. It is easy to realize the practical importance, for instance, of a
radio transmission carried out by localized beams, independently of their
being sub- or Super-luminal. \ But non-dispersive wave packets can be of use
also in theoretical physics for a reasonable representation of elementary
particles[A33].

\hspace*{5ex} Within Extended Relativity since 1980 it had been found[A34]
that ---whilst the simplest subluminal object conceivable is a small sphere,
or a point as its limit--- the simplest Superluminal objects results by
contrast to be (see refs.[A34], and Figs.\,A5 and A6) an ``X-shaped'' wave, or a
double cone as its limit, which moreover travels without deforming ---i.e.,
rigidly--- in a homogeneous medium[A3]. \ It is worth noticing that the most
interesting localized solutions happened to be just the Superluminal ones,
and with a shape of that kind. \ Even more, since from Maxwell equations
under simple hypotheses one goes on to the usual {\em scalar} wave equation
for each electric or magnetic field component, one can expect the same
solutions to exist also in the field of acoustic waves, and of seismic waves
(and perhaps of gravitational waves too). \ Actually, such beams (as
suitable superpositions of Bessel beams) were mathematically constructed for
the first time, {\em in acoustics\/}, by Lu et al.[A35], : and were then called
``X-waves'' or rather X-shaped waves.

\hspace*{5ex} It is more important for us that the X-shaped waves have been
in effect produced in experiments both with acoustic and with
electromagnetic waves; that is, X-beams were produced that, in their medium,
travel undistorted with a speed larger than sound, in the first case, and
than light, in the second case. \ In Acoustics, the first experiment was
performed by Lu et al. themselves[A36] in 1992, at the Mayo Clinic (and their
papers received the first 1992 IEEE award). \ In the {\em electromagnetic case},
certainly more ``intriguing'', Superluminal localized X-shaped solutions
were first mathematically constructed (cf., e.g., Fig.A7) in refs.[A37], and
later on experimentally produced by Saari et al.[A38] in 1997 at Tartu by
visible light (Fig.A8), and more recently by Mugnai, Ranfagni and Ruggeri at
Florence by microwaves[A39]. \ Further experimental activity is in progress, 
while in the theoretical sector the
activity is even more intense: in order to build up ---for example--- new
analogous solutions with finite total energy or more suitable for high
frequencies[A40], and localized solutions Superluminally propagating
even along a normal waveguide[A41], or in dispersive media[A42]; without
forgetting the aim of focusing X-shaped waves at a given space-point[A43],
and so on.

\hspace*{5ex} Let us eventually touch the problem of producing an X-shaped
Superluminal wave like the one in Fig.A6, but truncated ---of course-- in
space and in time (by the use of a finite, dynamic antenna, radiating for a
finite time): in such a situation, the wave will keep its localization and
Superluminality only along a certain ``depth of field'', decaying abruptly
afterwards[A35,A37,A44]. \ We can become convinced about the possibility of
realizing it, by imaging the simple ideal case of a negligibly sized
Superluminal source $S$ endowed with speed $V>c$ in vacuum and emitting
electromagnetic waves $W$ (each one travelling with the invariant speed $c$%
). The electromagnetic waves will result (cf. Fig.A6) to be internally tangent to an
enveloping cone $C$ having $S$ as its vertex, and as its axis the
propagation line $x$ of the source[A45,A34,A3]. \ This is analogous to what happens
for a plane that moves in the air with constant supersonic speed. \ The
waves $W$ interfere negatively inside the cone $C$, and constructively only
on its surface. \ We can place a plane detector orthogonally to $x$, and
record magnitude and direction of the $W$ waves that hit on it, as
(cylindrically symmetric) functions of position and of time. \ It will be
enough, then, to replace the plane detector with a plane antenna which {\em %
emits} ---instead of recording--- exactly the same (axially symmetric)
space-time pattern of waves $W$, for constructing a cone-shaped
electromagnetic wave $C$ that will propagate with the Superluminal speed $V$
(of course, without a source any longer at its vertex): \ even if each wave $%
W$ travels with the invariant speed $c$. \ For further details, see the
first one of refs.[A37], and refs.[A45,A34]. \ Here let us only add that such localized
Superluminal waves appear to keep their good properties only as long as they
are fed by the waves arriving (with speed $c$) from the dynamic antenna:
taking the time needed for their generation into account, these waves seem
therefore unable to transmit information faster than $c$; however, they have
nothing to do with the illusory ``scissors effect'', since they certainly
carry energy-momentum Superluminally along their field depth (for instance,
they can get two detectors at a distance $L$ to click after a time {\em %
smaller} than $L/c$). And some authors started recently considering the possibility
that the Superluminal localized solutions to the wave equations be actually
tachyonic: see. refs.[A46,A14], as well as the older refs.[A34,A6,A3].

\hspace*{5ex} We cannot end without calling attention to some recent,
interesting experimental results, presented in refs.[A47], which regard
the production of X-shaped waves in {\em non-linear materials}.

\hspace*{5ex} As we mentioned above, the existence of all these X-shaped
Superluminal (or ``Super-sonic'') seem to constitute at the
moment---together, e.g., with the Superluminality of evanescent waves--- one
confirmation of Extended Relativity. \ But, at present, the existence of
localized (non-diffractive) pulses or wave-trains, even if rather interesting
for theory, appear to be even more important for their possible
{\em applications\/}: One of the first applications of such X-waves
(that takes advantage just of their propagation without deformation) is in
advanced progress in the field of medicine, and precisely of ultrasound
scanners[A48].

\newpage

{\bf Captions of the Figures of the Appendix}

\

Fig.A1 -- Energy of a free object as a function of its speed[A2-A4].\hfill\break

Fig.A2 -- Depicting the ``switching rule" (or reinterpretation principle) by
Stueckelberg-Feynman-Sudarshan-Recami[A3-A5]: particle $Q$ will appear as the
antiparticle of P. \ See the text.\hfill\break

Fig.A3 -- Simulation of tunnelling by experiments with classical evanescent 
waves (see the text), which were predicted to be Superluminal also on the
basis of Extended Relativity[A3-A4]. The figure shows one of the experimental
results in refs.[15]: that is, the average speed of the wave while crossing
the evanescence region, as a function of its length.
In the present case, the classical ``barrier" is a segment of undersized
waveguide. \ As theoretically predicted[A19,A12], such an
average speed exceeds $c$ for long enough barriers. (In this figure,
the barrier width has been called $a$, instead of $d$, at variance with our
choices).\hfill\break

Fig.A4 -- The very interesting experiment having recourse to a metallic
waveguide with TWO ``barriers" (undersized guide segments), i.e., with two
evanescence regions[A22]. See the text.\hfill\break

Fig.A5 -- An intrinsically spherical (or pointlike, in the limiting case)
object appears in the vacuum as an ellipsoid contracted along the motion
direction ---due to Lorentz contraction--- when endowed with a speed $v<c$. \
By contrast, if endowed with a speed
$V>c$ (even if the $c$-speed barrier cannot be crossed, neither from the
left nor from the right), it would appear no longer as a particle, but
rather as an ``X-shaped" wave[A34] travelling rigidly (namely, as occupying
the region delimited by a double cone and a two-sheeted hyperboloid ---or
as a double cone, in the considered limiting case--, moving Superluminally
and without distortion in the vacuum, or in a homogeneous medium).\hfill\break

Fig.A6 -- Here the intersections are shown of an ``X-shaped wave"[A34] with
planes orthogonal to its motion line, according to Extended
Relativity``[A2-A4].
The examination of this figure suggests how to construct a simple dynamic
antenna for generating such localized Superluminal waves (such an antenna
was in fact adopted, independently, by Lu et al.[A36] for the production
of non-diffractive beams)\hfill\break

Fig.A7 -- Theoretical prediction of the Superluminal localized ``X-shaped"
waves for the electromagnetic case (from Lu, Greenleaf and Recami[A37],
and Recami[A37]).\hfill\break

Fig.A8 -- Scheme of the experiment by Saari et al.[A38], who announced
(PRL of 24 Nov.1997) the production in optics of the beams depicted in Fig.8:
In this
figure one can see what shown by the experiment: i.e., the Superluminal
``X-shaped" waves, which run after and catch up with the plane waves (the
latter regularly travelling with speed $c$). An analogous experiment has
been later on (PRL of 22 May 2000) performed with microwaves at Florence
by Mugnai, Ranfagni and Ruggeri[A39].

\newpage

{\bf REFERENCES of the Appendix}\hfill\break

[A1] See, e.g., O.M.Bilaniuk, V.K.Deshpande \& E.C.G.Sudarshan: Am. J. Phys.
30 (1962) 718.

[A2] See E.Recami \& R.Mignani: Rivista N. Cim. 4 (1974) 209-290; 4 (1974)
E398, and refs. therein. \ Cf. also E.Recami (editor): {\em Tachyons,
Monopoles, and Related Topics} (North-Holland; Amsterdam, 1978).

[A3] E.Recami: Rivista N. Cim. 9 (1986), issue no.6, pp.1$\div$178, and
refs. therein.

[A4] See, e.g., E.Recami: in {\em Annuario 73, Enciclopedia EST}, ed. by
E.Macorini (Mondadori; Milano, 1973), pp.85-94; \ and \ Nuovo Saggiatore 2
(1986), issue no.3, pp.20-29.

[A5] E.Recami: in {\em I Concetti della Fisica}, ed. by F.Pollini \&
G.Tarozzi (Acc.Naz.Sc.Lett.Arti; Modena, 1993), pp.125-138; \ E.Recami \&
W.A.Rodrigues: ``Antiparticles from Special Relativity", Found. Physics
12 (1982) 709-718; 13 (1983) E533.

[A6] E.Recami: Found. Physics 17 (1987) 239-296. \ See also Lett. Nuovo
Cimento 44 (1985) 587-593; \ and P.Caldirola \& E.Recami: in {\em Italian
Studies in the Philosophy of Science}, ed. by M.Dalla Chiara (Reidel;
Boston, 1980), pp.249-298.

[A7] Cf. M.Baldo Ceolin: ``Review of neutrino physics", invited talk at the
{\em XXIII Int. Symp. on Multiparticle Dynamics (Aspen, CO; Sept.1993)}; \
E.W.Otten: Nucl. Phys. News 5 (1995) 11. \ From the theoretical point of
view, see, e.g., E.Giannetto, G.D. Maccarrone, R.Mignani \& E.Recami: Phys.
Lett. B178 (1986) 115-120 and refs. therein; \ S.Giani: ``Experimental
evidence of Superluminal velocities in astrophysics and proposed
experiments", CP458, in {\em Space Technology and Applications International
Forum 1999}, ed. by M.S.El-Genk (A.I.P.; Melville, 1999), pp.881-888.

[A8] See, e.g., J.A.Zensus \& T.J.Pearson (editors): {\em Superluminal Radio
Sources} (Cambridge Univ.Press; Cambridge, UK, 1987).

[A9] I.F.Mirabel \& L.F.Rodriguez. : ``A Superluminal source in the Galaxy",
Nature 371 (1994) 46 [with an editorial comment, ``A galactic speed record",
by G.Gisler, at page 18 of the same issue]; \ S.J.Tingay et al.:
``Relativistic motion in a nearby bright X-ray source", Nature 374 (1995)
141.

[A10] M.J.Rees: Nature 211 (1966) 46; \ A.Cavaliere, P.Morrison \& L.Sartori:
Science 173 (1971) 525.

[A11] E.Recami, A.Castellino, G.D.Maccarrone \& M.Rodon\`o: ``Considerations
about the apparent Superluminal expansions observed in astrophysics", Nuovo
Cimento B93 (1986) 119. \ Cf. also R.Mignani \& E.Recami: Gen. Relat. Grav.
5 (1974) 615.

[A12] V.S.Olkhovsky \& E.Recami: Phys. Reports 214 (1992) 339, and refs.
therein: In particular T.E.Hartman: J. Appl. Phys. 33 (1962) 3427. \ See
also V.S.Olkhovsky, E.Recami, F.Raciti \& A.K.Zaichenko: J. de Phys.-I
5 (1995) 1351-1365; \ V.S.Olkhovsky, E.Recami \& J.Jakiel: ``Unified time analysis of
photon and nonrelativistic particle tunnelling" [e-print quant-ph/0102007],
to appear in Physics Reports.

[A13] See, e.g., Th.Martin \& R.Landauer: Phys. Rev. A45 (1992) 2611; \
R.Y.Chiao, P.G.Kwiat \& A.M.Steinberg: Physica B175 (1991) 257; \
A.Ranfagni, D.Mugnai, P.Fabeni \& G.P.Pazzi: Appl. Phys. Lett. 58 (1991)
774; \ Y.Japha \& G.Kurizki: Phys. Rev. A53 (1996) 586. \ Cf. also
G.Kurizki, A.E.Kozhekin \& A.G.Kofman: Europhys. Lett. 42 (1998) 499: \
G.Kurizki, A.E.Kozhekin, A.G.Kofman \& M.Blaauboer: paper delivered at the
VII Seminar on Quantum Optics, Raubichi, BELARUS (May, 1998).

[A14] E.Recami, F.Fontana \& R.Garavaglia: Int. J. Mod. Phys. A15 (2000)
2793, and refs. therein.

[A15] G.Nimtz \& A.Enders: J. de Physique-I 2 (1992) 1693; \ 3 (1993) 1089; \
4 (1994) 1379; \ Phys. Rev. E48 (1993) 632; \ H.M.Brodowsky, W.Heitmann \&
G.Nimtz: J. de Physique-I 4 (1994) 565; \ Phys. Lett. A222 (1996) 125; A196
(1994) 154; \ G.Nimtz and W.Heitmann: Prog. Quant. Electr. 21 (1997) 81.

[A16] A.M.Steinberg, P.G.Kwiat \& R.Y.Chiao: Phys. Rev. Lett. 71 (1993) 708,
and refs. therein; \ Scient. Am. 269 (1993) issue no.2, p.38. \ Cf. also
Y.Japha \& G.Kurizki: Phys. Rev. A53 (1996) 586.

[A17] A.Ranfagni, P.Fabeni, G.P.Pazzi \& D.Mugnai: Phys. Rev. E48 (1993)
1453; \ Ch.Spielmann, R.Szipocs, A.Stingl \& F.Krausz: Phys. Rev. Lett. 73
(1994) 2308, \ Ph.Balcou \& L.Dutriaux: Phys. Rev. Lett. 78 (1997) 851; \
V.Laude \& P.Tournois: J. Opt. Soc. Am. B16 (1999) 194.

[A18] Scientific American (Aug. 1993);\ Nature (Oct.21, 1993); \ New
Scientist (Apr. 1995); \ Newsweek (19 June 1995).

[A19] Ref.[A3], p.158 and pp.116-117. \ Cf. also D.Mugnai, A.Ranfagni,
R.Ruggeri, A.Agresti \& E.Recami: Phys. Lett. A209 (1995) 227.

[A20] H.M.Brodowsky, W.Heitmann \& G.Nimtz: Phys. Lett. A222 (1996) 125.

[A21] A.P.L.Barbero, H.E.Hern\'andez F., \& E.Recami: ``On the propagation
speed of evanescent modes" [e-print physics/9811001], Phys. Rev. E62
(2000) 8628-8635, and refs. therein. \ See also E.Recami, H.E.Hern\'andez F.,
\& A.P.L.Barbero: Ann. der Phys. 7 (1998) 764-773; \ and, for instance, \
R.Y.Chiao and
A.M.Steinberg: in {\em Progress in Optics}, ed.by E.Wolf (Elsevier Science;
Amsterdam, 1997), vol.37, pp.346-405; \ G.Nimtz and W.Heitman: Progr. Quant.
Electr. 21 (1997) 81-108; \ R.W.Ziolkowski: Phys. Rev. E63 (2001) 046604; \
A.M.Shaarawi and I.M.Besieris: J. Phys. A:Math.Gen. 33 (2000) 7227-7254;
7255-7263; \ P.W.Milonni: J. Phys. B35 (2002) R31-R56; \ G.Nimtz: IEEE J.
Select. Top. Quantum Electron. 9(1) (2003) 163-171.

[A22] G.Nimtz, A.Enders \& H.Spieker: in {\em Wave and Particle in Light and
Matter}, ed. by A.van der Merwe \& A.Garuccio (Plenum; New York, 1993); \ J.
de Physique-I 4 (1994) 565. \ See also A.Enders \& G.Nimtz: Phys. Rev. B47
(1993) 9605.

[A23] V.S.Olkhovsky, R.Recami \& G.Salesi: Europhy. Lett. 57,
879 (2002). \ See also Y. Aharonov, N. Erez \& B. Reznik: Phys. Rev. A65 (2002)
no.052124; \ S. Longhi, P. Laporta, M. Belmonte \& E. Recami:``Measurement
of Superluminal optical tunneling times in double-barrier photonic bandgaps",
Phys. Rev. E65 (2002) no.046610; \ and \ S. Esposito: Phys. Rev.
E67 (2003) no.016609.

[A24] V.S.Olkhovsky, E.Recami, F.Raciti \& A.K.Zaichenko: ref.[A12], p.1361.
\ See also refs.[A3,A6], and E.Recami, F.Fontana \& R.Garavaglia: ref.[A14],
p.2807.

[A25] R.Y.Chiao, A.E.Kozhekin A.E., and G.Kurizki: Phys. Rev. Lett. 77 (1996)
1254; \ C.G.B.Garret \& D.E.McCumber: Phys. Rev. A1 (1970) 305.

[A26] S.Chu \& Wong W.: Phys. Rev. Lett. 48 (1982) 738; \ M.W.Mitchell \&
R.Y.Chiao: Phys. Lett. A230 (1997) 133-138; \ G.Nimtz: Europ. Phys. J. B (to
appear as a Rapid Note); \ L.J.Wang, A.Kuzmich \& A.Dogariu: Nature 406
(2000) 277.

[A27] G.Alzetta, A.Gozzini, L.Moi \& G.Orriols: Nuovo Cimento B36B (1976) 5.

[A28] M.Artoni, G.C.La Rocca, F.S. Cataliotti \& F. Bassani: Phys. Rev. A (in
press).

[A29] H.Bateman: {\em Electrical and Optical Wave Motion} (Cambridge
Univ.Press; Cambridge, 1915); \ R.Courant \& D.Hilbert: {\em Methods of
Mathematical Physics} (J.Wiley; New York, 1966), vol.2, p.760; \
J.N.Brittingham: J. Appl. Phys. 54 (1983) 1179; \ R.W.Ziolkowski: J. Math.
Phys. 26 (1985) 861; \ J.Durnin: J. Opt. Soc. 4 (1987) 651; \ A.O.Barut et
al.: Phys. Lett. A143 (1990) 349; \ Found. Phys. Lett. 3 (1990) 303; \
Found. Phys. 22 (1992) 1267.

[A30] J.A.Stratton: {\em Electromagnetic Theory} (McGraw-Hill; New York,
1941), p.356; \ A.O.Barut et al.: Phys. Lett. A180 (1993) 5; A189 (1994) 277.

[A31] R.Donnelly \& R.W.Ziolkowski: Proc. Roy. Soc. London A440 (1993) 541; \
I.M.Besieris, A.M.Shaarawi \& R.W.Ziolkowski: J. Math. Phys. 30 (1989) 1254;
\ S.Esposito: Phys. Lett. A225 (1997) 203; \ J.Vaz \& W.A.Rodrigues; Adv.
Appl. Cliff. Alg. S-7 (1997) 457.

[A32] See also E.Recami \& W.A.Rodrigues Jr.: ``A model theory for tachyons
in two dimensions", in {\em Gravitational Radiation and Relativity}, ed. by
J.Weber \& T.M.Karade (World Scient.; Singapore, 1985), pp.151-203, and
refs. therein.

[A33] A.M.Shaarawi, I.M.Besieris and R.W.Ziolkowski: J. Math. Phys. 31 (1990)
2511, Sect.VI; \ Nucl Phys. (Proc.Suppl.) B6 (1989) 255; \ Phys. Lett. A188
(1994) 218. \ See also V.K.Ignatovich: Found. Phys. 8 (1978) 565; and
A.O.Barut: Phys. Lett. A171 (1992) 1; A189 (1994) 277; Ann. Foundation L. de
Broglie, Jan.1994; and ``Quantum theory of single events: Localized de
Broglie--wavelets, Schroedinger waves and classical trajectories", preprint
IC/90/99 (ICTP; Trieste, 1990).

[A34] A.O.Barut, G.D.Maccarrone \& E.Recami: Nuovo Cimento A71 (1982) 509; \
P.Caldirola, G.D.Maccarrone \& E.Recami: Lett. Nuovo Cim. 29 (1980) 241; \
E.Recami \& G.D.Maccarrone: Lett. Nuovo Cim. 28 (1980) 151.

[A35] J.-y.Lu \& J.F.Greenleaf: IEEE Trans. Ultrason. Ferroelectr. Freq.
Control 39 (1992) 19.

[A36] J.-y.Lu \& J.F.Greenleaf: IEEE Trans. Ultrason. Ferroelectr. Freq.
Control 39 (1992) 441.

[A37] E.Recami: Physica A252 (1998) 586; \ J.-y.Lu, J.F.Greenleaf \&
E.Recami: ``Limited diffraction solutions to Maxwell (and Schroedinger)
equations'' [e-print physics/9610012], Report INFN/FM--96/01
(I.N.F.N.; Frascati, Oct.1996); \ R.W.Ziolkowski, I.M.Besieris \&
A.M.Shaarawi: J. Opt. Soc. Am. A10 (1993) 75.

[A38] P.Saari \& K.Reivelt: ``Evidence of X-shaped propagation-invariant
localized light waves", Phys. Rev. Lett. 79 (1997) 4135-4138.

[A39] D.Mugnai, A.Ranfagni \& R.Ruggeri: Phys. Rev. Lett. 84 (2000)
4830.

[A40] See M.Z.Rached, E.Recami \& H.E.Hern\'{a}ndez-Figueroa:
Europ. Phys. Journal D21 (2002) 217-228, and refs. therein. \ See also
I.M.Besieris, M.Abdel-Rahman, A.Shaarawi \& A.Chatzipetros: Progress in
Electromagnetic Research (PIER) 19 (1998) 1-48; \ A.T.Friberg, J.Fagerholm
\& M.M.Salomaa: Opt. Commun. 136 (1997) 207; \ J.Fagerholm, A.T.Friberg,
J.Huttunen, D.P.Morgan \& M.M.Salomaa: Phys. Rev. E54 (1996) 4347; \
A.M.Shaarawi: J. Opt. Soc. Am. A14 (1997) 1804-1816; \ A.M.Shaarawi,
I.M.Besieris, R.W.Ziolkowski \& R.M.Sedky: J. Opt. Soc. Am. A12 (1995)
1954-1964; \ P.Saari: in {\it Time's Arrows, Quantum Measurements
and Superluminal Behavior}, ed. by D.Mugnai et al. (C.N.R.; Rome, 2001),
pp.37-48.

[A41] M.Z.Rached, E.Recami \& F.Fontana: ``Localized Superluminal
solutions to Maxwell equations propagating along a normal-sized waveguide'',
Phys. Rev. E64 (2001) no.066603; \ M.Z.Rached, F.Fontana \& E.Recami:
``Superluminal localized solutions
to Maxwell equations along a waveguide: The finite-energy case",
Phys. Rev. E67 (2003) 036620 [7 pages]; \ M.Z.Rached,
K.Z.N\'obrega, E.Recami \& H.E.Hern\'andez F.: ``Superluminal
X-shaped beams propagating without distortion along a coaxial guide",
Phys. Rev. E66 (2002) 046617 [10 pages]. \
Cf. also, e.g., M.A.Porras, R.Borghi \& M.Santarsiero: ``Superluminality in
Gaussian beams", Opt. Commun. 203 (2002) 183-189.

[A42] See M.Z.Rached, K.Z.N\'obrega, H.E.Hern\'{a}ndez-Figueroa \&
E.Recami: ``Localized Superluminal solutions to the wave equation in
(vacuum or) dispersive media, for arbitrary frequencies and with
adjustable bandwidth" [e-print physics/0209101],
Opt. Commun. 226 (2003) 15-23, and refs. therein; \ P.Saari \& H.S\~{o}najalg: ``Pulsed Bessel beams",
Laser Phys. 7 (1997) 32-39. \ See also E.Recami, M.Z.Rached, K.Z.N\'obrega,
C.A.Dartora \& H.E.Hern\'andez F.: ``On the localized Superluminal solutions
to the Maxwell equations", IEEE Journal of Selected Topics in Quantum
Electronics 9(1) (2003) 59-73, which is in part a (brief) review
article.

[A43] See M.Z.Rached, A.M.Shaarawi \& E.Recami: ``Focused X-shaped
(Superluminal) pulses" [e-print physics/0309098], submitted for pub.,
and refs. therein; \ A.M.Shaarawi, I.M.Besieris \& T.M.Said:
``Temporal focusing by use of composite X-waves", J. Opt. Soc. Am.
A20 (2003) 1658-1665.

[A44] See, e.g., C.A.Dartora, M.Z.Rached \& E.Recami: ``General formulation
for the analysis of scalar diffraction-free beams using angular modulation:
Mathieu and Bessel beams", Opt. Commun. 222 (2003) 75-85, and refs.
therein.

[A45]  E.Recami, M.Zamboni Rached \& C.A.Dartora: ``The X-shaped, localized
field generated by a Superluminal electric charge",
accepted for pub. in Phys. Rev. E as a Brief Report.\\

[A46] R.W.Ziolkowski \& A.D.Kipple: ``Causality and double-negative
metamaterials", Phys. Rev. E69 (2003) 026615 [9 pages]; \ E.Recami:
``Superluminal motions? A bird-eye view of the experimental
status-of-the-art" [e-print physics/0101108], Found. Phys. 31 (2001)
1119-1135. \ See also E.Recami, M.Z.Rached, K.Z.N\'obrega,
C.A.Dartora \& H.E.Hern\'andez F.: ref.[A42].

[A47] See C.Conti, S.Trillo, P.Di Trapani, G.Valiulis, A.Piskarskas,
O.Jedrkiewicz \& J.Trull: ``Nonlinear electromagnetic X-waves",
Phys. Rev. Lett. 90 (2003) 170406 [4 pages]. \
Cf. also M.A.Porras, S.Trillo, C.Conti and P.Di Trapani:
``Paraxial envelope X-waves", Opt. Lett. 28 (2003) 1090-1092.

[A48] J.-y.Lu, H.-h.Zou \& J.F.Greenleaf: Ultrasound in Medicine and Biology
20 (1994) 403; \ Ultrasonic Imaging 15 (1993) 134.

\end{document}